\documentclass[twocolumn,twocolappendix,times]{aastex63}

\usepackage{savesym}
\savesymbol{splitbox}
\usepackage[export]{adjustbox}
\restoresymbol{TXF}{splitbox}
\usepackage{multirow}   
\usepackage{amsmath}
\usepackage{CJK}

\usepackage{hyperref}
\hypersetup{
    colorlinks=true,
    linkcolor=blue,
    filecolor=blue,      
    urlcolor=blue,
}
\newcommand{\xray}{\hbox{X-ray}}  

\newcommand{\coii}{CO(2--1)}
\newcommand{\coiii}{CO(3--2)}

\newcommand{\mbg}{M_{\rm bulge}} 
\newcommand{\mstar}{M_{\rm star}} 
\newcommand{\mbh}{M_{\rm BH}} 
\newcommand{\mhh}{M_{\rm gas}}
\newcommand{\mdust}{M_{\rm dust}}

\newcommand{\lpco}{L'_{\rm CO}}
\newcommand{\lef}{L_{\rm \nu,850\ \mu m}}
\newcommand{\ico}{I_{\rm CO}}
\newcommand{\aco}{\alpha_{\rm CO}}
\newcommand{\fracA}{{\rm frac_{AGN}}}
\newcommand{\ox}{\alpha_{\rm ox}}
\newcommand{\luvr}{L_{\nu, 2500 \textup{\AA}}}
\newcommand{\avism}{A_{V}^{\rm ISM}}
\newcommand{\avbc}{A_{V}^{\rm BC}}
\newcommand{\lgalir}{L_{\rm IR, gal}}
\newcommand{\tdep}{\tau_{\rm dep}}

\newcommand{\tua}{t_{\rm ua}}
\newcommand{\tgrow}{\tau_{\rm grow}}
\newcommand{\tgbh}{\tau_{\rm grow, BH}}

\newcommand{\cig}{{\sc cigale}}
\newcommand{\casa}{{\sc casa}}

\newcommand{\spitzer}{{\it Spitzer\/}}
\newcommand{\herschel}{{\it Herschel\/}}

\newcommand{\fst}[1]{#1}
\newcommand{\fstm}[1]{#1}
\newcommand{\scd}[1]{#1}

\shorttitle{gas in bulges}
\shortauthors{Yang et al.}

\begin{document}
\begin{CJK*}{UTF8}{gbsn}

\title{Does the lockstep growth between black holes and bulges create their mass relation?}

\correspondingauthor{Guang Yang}
\email{gyang206265@gmail.com}

\author{Guang Yang (杨光)}
\affiliation{Department of Physics and Astronomy, Texas A\&M University, College Station, TX 77843-4242, USA}
\affiliation{George P. and Cynthia Woods Mitchell Institute for Fundamental Physics and Astronomy, Texas A\&M University, College Station, TX 77843-4242, USA}
\affiliation{Kapteyn Astronomical Institute, University of Groningen, P.O. Box 800, 9700 AV Groningen, The Netherlands}
\affiliation{SRON Netherlands Institute for Space Research, Postbus 800, 9700 AV Groningen, The Netherlands}

\author{W. N. Brandt}
\affiliation{Department of Astronomy and Astrophysics, 525 Davey Lab, The Pennsylvania State University, University Park, PA 16802, USA}
\affiliation{Institute for Gravitation and the Cosmos, The Pennsylvania State University, University Park, PA 16802, USA}
\affiliation{Department of Physics, 104 Davey Laboratory, The Pennsylvania State University, University Park, PA 16802, USA}

\author{David M. Alexander}
\affiliation{Centre for Extragalactic Astronomy, Department of Physics, Durham University, South Road, Durham DH1 3LE, UK}

\author{M\'ed\'eric Boquien}
\affiliation{Centro de Astronom\'ia (CITEVA), Universidad de Antofagasta, Avenida Angamos 601, Antofagasta, Chile}

\author{Qingling Ni}
\affiliation{Institute for Astronomy, University of Edinburgh, Royal Observatory, Edinburgh, EH9 3HJ, UK}

\author{Casey Papovich}
\affiliation{Department of Physics and Astronomy, Texas A\&M University, College Station, TX 77843-4242, USA}
\affiliation{George P. and Cynthia Woods Mitchell Institute for Fundamental Physics and Astronomy, Texas A\&M University, College Station, TX 77843-4242, USA}

\author{Justin S. Spilker}
\affiliation{Department of Physics and Astronomy, Texas A\&M University, College Station, TX 77843-4242, USA}
\affiliation{George P. and Cynthia Woods Mitchell Institute for Fundamental Physics and Astronomy, Texas A\&M University, College Station, TX 77843-4242, USA}

\author{Fabio Vito}
\affiliation{Scuola Normale Superiore, Piazza dei Cavalieri 7, I-56126 Pisa, Italy}

\author{Jonelle L. Walsh}
\affiliation{Department of Physics and Astronomy, Texas A\&M University, College Station, TX 77843-4242, USA}
\affiliation{George P. and Cynthia Woods Mitchell Institute for Fundamental Physics and Astronomy, Texas A\&M University, College Station, TX 77843-4242, USA}

\author{Chengpeng Zhang}
\affiliation{Department of Physics and Astronomy, Texas A\&M University, College Station, TX 77843-4242, USA}
\affiliation{George P. and Cynthia Woods Mitchell Institute for Fundamental Physics and Astronomy, Texas A\&M University, College Station, TX 77843-4242, USA}



\begin{abstract}

Recent studies have revealed a strong relation between sample-averaged black-hole (BH) accretion rate (BHAR) and star formation rate (SFR) among bulge-dominated galaxies, i.e., ``lockstep'' BH-bulge growth, in the distant universe. 
This relation might be closely related to the BH-bulge mass correlation observed in the local universe. 
To understand further BH-bulge coevolution, we present ALMA \coii\ or \coiii\ observations of 7 star-forming bulge-dominated galaxies at $z=0.5$--2.5.
Using the ALMA data, we detect significant ($>3\sigma$) CO emission from 4 objects. 
For our sample of 7 galaxies, we measure (or constrain with upper limits) their CO line fluxes and estimate molecular gas masses ($\mhh$).
We also estimate their stellar masses ($\mstar$) and SFRs by modelling their spectral energy distributions (SEDs).
Using these physical properties, we derive the gas-depletion timescales ($\tdep \equiv \mhh/{\rm SFR}$) and compare them with the bulge/BH growth timescales ($\tgrow \equiv \mstar/{\rm SFR} \sim \mbh/{\rm BHAR}$).
Our sample generally has $\tdep$ shorter than $\tgrow$ by a median factor of $\gtrsim 4$, indicating that the cold gas will be depleted before significant bulge/BH growth takes place.
This result suggests that the BH-bulge lockstep growth is mainly responsible for maintaining their mass relation, not creating it.
We note that our sample is small and limited to $z<2.5$;
\textit{JWST}\ and ALMA will be able to probe to higher redshifts in the near future. 

\end{abstract}



\section{Introduction} 
\label{sec:intro}
From observations of the nearby universe, massive galaxies always host supermassive black holes (BHs) in their central regions, and the BH mass ($\mbh$) is tightly correlated with the stellar mass ($\mstar$) of the host-galaxy bulge \citep[e.g.,][]{kormendy13, saglia16}.
This tight BH-bulge mass correlation indicates that a physical connection between the BHs and bulges exists in their cosmic evolution history.
This connection is often called ``BH-bulge coevolution''.

The specific coevolution mechanisms remain largely unknown.
Some early studies proposed minor-merger events as the driver of the tight BH-bulge mass relation in the local universe \citep[e.g.,][]{peng07, jahnke11}.
The idea is based on the statistical central-limit theorem: if low-mass galaxies are scattered around the $\mbh$-$\mstar$ relation, the final system will be close to the mass relation after many episodes of minor mergers. 
This scheme requires that low-mass systems are centered around the $\mbh$-$\mstar$ relation.
However, more recent observations indicate that low-mass galaxies have $\mbh$ systematically below the relation \citep[e.g.,][]{aird18, yang18} and that some low-mass galaxies may not even  host BHs \citep[e.g.,][]{greene20}. 
Therefore, minor mergers are unlikely to be the origin of the BH-bulge mass relation. 
\fst{Another idea about the origin of the $\mbh$-$\mstar$ relation is related to AGN feedback.
A BH grows to a critical mass (related to $\mbg$) and launches a powerful wind that removes cold gas and/or prevents gas replenishment \citep[e.g.,][]{king15}. 
The BH and stellar growth are thereby halted due to the lack of fuel.
This scenario requires strong AGN negative feedback on star formation. 
However, this assumption still lacks strong observational support \citep[e.g.,][]{harrison17, shangguan20}.}

\fst{From an observational point of view, the $\mbh$-$\mstar$ relation might result from some connections between BH accretion and host galaxies.}
Earlier observations on this topic mostly focused on the connections of BH growth versus host-galaxy $\mstar$ and star formation rate (SFR) at different redshifts (see \S3.3 of \citealt{brandt21} for a review).
Some significant relations have been found.
For example, BH growth strongly depends upon $\mstar$ at a given redshift \citep[e.g.,][]{georgakakis17, yang17, yang18, aird18}.
However, it is not obvious how these relations are related to the BH-bulge correlation in the local universe, mainly because these relations are for global galaxy (rather than bulge) properties.
It is technically challenging to investigate bulges, since galaxy angular sizes are generally small ($\lesssim$~arcsec scales) in the distant universe ($z\gtrsim 1$).
Also, to avoid strong effects from the ``morphological K correction'' in the rest-frame UV, the sampled rest-frame wavelengths are required to be longer than $\approx 4000 \textup{\AA}$ \citep[e.g.,][]{papovich05, conselice14}.
This means that the imaging should use IR bands ($\gtrsim 1.2\ \mu$m) to measure the morphologies of sources at $z\approx 2$, the cosmic epoch when active galactic nucleus (AGN) and star formation (SF) activities peak.
The Cosmic Assembly Near-infrared Deep Extragalactic Legacy Survey (CANDELS; \citealt{grogin11,koekemoer11}) provides high-angular-resolution ($\approx 0.2''$) and deep $H$-band imaging over a $\approx 900$~arcmin$^2$ area, providing an excellent chance to study AGN-bulge connections at $z\lesssim 3$.

Using the CANDELS-based morphological measurements \citep{huertas_company15} as well as other multiwavelength data, \cite{yang19} selected a sample of bulge-dominated galaxies and investigated their sample-averaged BH accretion rate (BHAR) versus SFR.\footnote{\label{ft:avg}This sample-averaged BHAR is designed to overcome AGN short-term ($\lesssim$~10$^7$~years) variability effects and approximate long-term BH growth rate \citep[e.g.,][]{hickox14}.
Therefore, individual bulge-dominated systems should follow the same BHAR-SFR relation over cosmic evolution timescales ($\gtrsim 10^{8}$ years).} 
They found a significant linear correlation between BHAR and SFR for the bulge-dominated galaxies, and this correlation does not exist among their comparison sample that has disks and/or irregularities (see, e.g., \citealt{kocevski17, ni19, ni21} for related results).
The best-fit BHAR/SFR ratio in \cite{yang19} is $\approx$~1/300, similar to the BH/bulge mass ratio observed in the local universe \citep[e.g.,][]{kormendy13}.
This ``lockstep'' style growth between BHs and bulges can be useful for predicting BH accretion from bulge SF information. 
Based on a sample of bulge-dominated galaxies with well-measured star formation histories (SFHs; \citealt{estrada_carpenter20}),  \cite{yang21b} predicted BH accretion densities at high redshifts, which can be tested with \textit{JWST} and future IR missions. 

\begin{figure*}[ht]
    \centering
	\includegraphics[width=2\columnwidth]{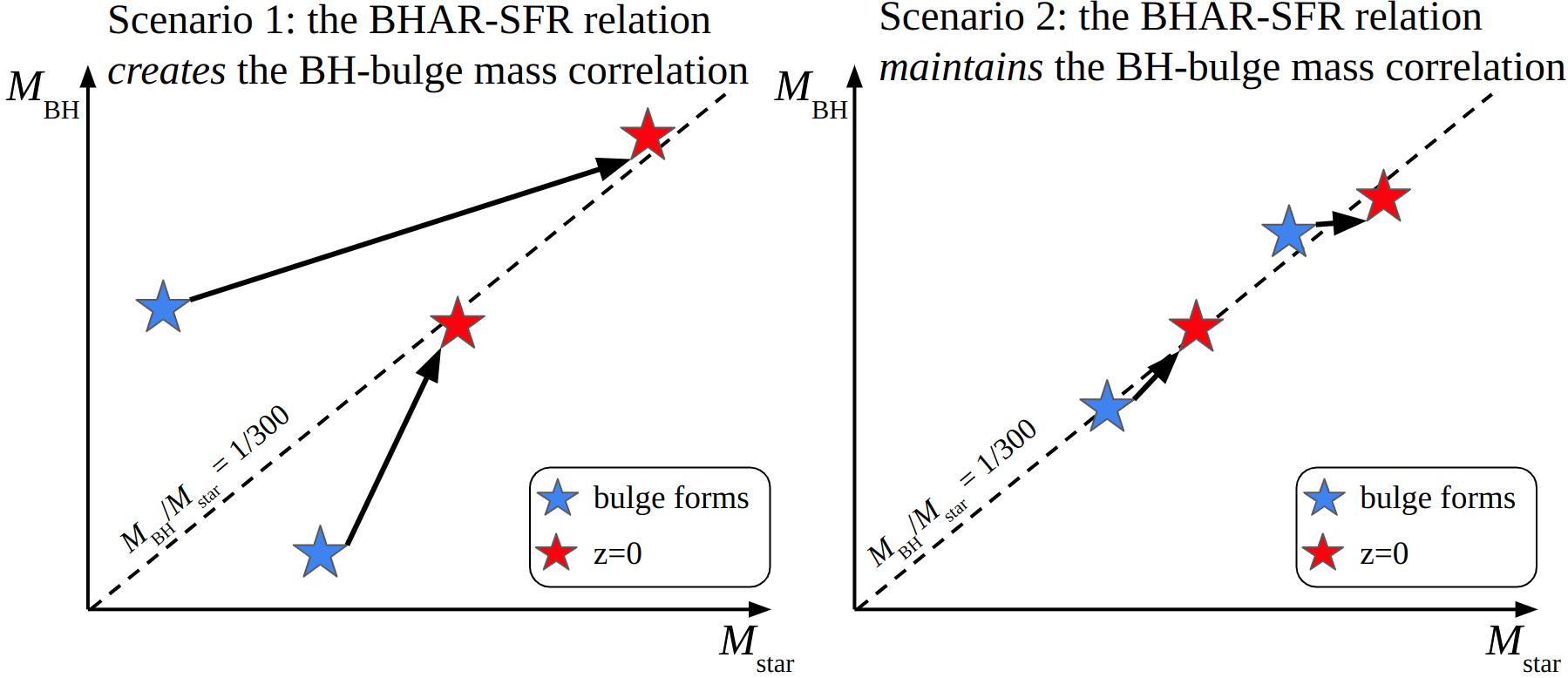}
    \caption{Schematic diagram of the two scenarios tested in this work. 
    The blue and red stars represent the bulge-formation redshift and $z=0$, respectively. 
    The black arrows indicate the evolution paths.
    The dashed line represents the observed BH-bulge mass relation in the local universe.  
    The determining factor between the two scenarios is the amount of cold gas available after bulge formation (see \S\ref{sec:intro} for details). 
    }
    \label{fig:scheme}
\end{figure*}

The BHAR-bulge SFR relation could be closely related to the BH-bulge mass relation in the local universe, given their similarities as above.  
However, the detailed picture of BH-bulge coevolution is still not clear, and there are two possible scenarios following the formation of the bulge: 
\begin{enumerate}

    \item  The system initially retains a large amount of cold gas, which is available to fuel significant bulge growth. 
    The BH then would continue to grow following the BHAR-SFR relation. 
    The galaxy thereby moves to $\mbh/\mstar \approx \rm BHAR/SFR \approx 1/300$, regardless of the  previous ratio of $\mbh/\mstar$ in the pre-bulge phase. 
    In this case, the BHAR-SFR relation plays a role of ``creation'' for the BH-bulge mass correlation. 
    
    \item The system initially has a limited or insignificant amount of cold gas remaining.  Star formation in the bulge will soon shut down soon due to the lack of fuel, \scd{assuming no further gas replenishment}. 
    The galaxy thereby remains in a low-specific SFR state and has a low-BHAR (according to the BHAR-SFR relation), and it passively evolves.
    The $\mbh/\mstar$ ``freezes'' at the value in the pre-bulge phase.
    In this case, the BHAR-SFR relation plays a role of ``maintenance'' for the BH-bulge mass correlation. 
    
\end{enumerate}
Fig.~\ref{fig:scheme} illustrates these two scenarios above schematically. 
The determining factor between the two scenarios is the amount of cold gas available after bulge formation. 
If gas is abundant, then the more realistic scenario is the first one. 
\fst{The system has significant BH/bulge mass growth, reducing the scatter around the local $\mbh$-$\mstar$ relation (Fig.~\ref{fig:scheme} left).
Otherwise, if gas is limited, the system does not have much fuel to grow its BH/bulge mass, and its position on the $\mbh$-$\mstar$ diagram remains roughly the same.
In this case, since the $z=0$ position is near the $\mbh$-$\mstar$ relation, the current position should also be near the relation (Fig.~\ref{fig:scheme} right).
}

In this work, we assess the gas content of bulge-dominated galaxies using ALMA observations.  
Our ALMA observation targeted \coii\ or \coiii\ for 7 bulge-dominated galaxies. 
For our targets, the low-$J$ transitions are the best-possible lines accessible by ALMA to estimate $\mhh$, as they are not subject to strong assumptions about the CO spectral line energy distribution (SLED).
Based on the ALMA data, we constrain the molecular-gas content and discuss the implications for BH-bulge coevolution. 
We note that each of the 7 individual systems should follow the BHAR-SFR relation (see Footnote~\ref{ft:avg}), and thus they are suitable to test the two scenarios laid out in Fig.~\ref{fig:scheme}.

The structure of this paper is as follows.
In \S\ref{sec:analyis}, we reduce and analyze the ALMA data. 
We also compile existing multiwavelength data and perform a spectral energy distribution (SED) analysis. 
In \S\ref{sec:discuss}, we discuss the physical implications of our results. 
We summarize our results and discuss future prospects in \S\ref{sec:sum}.

Throughout this paper, we assume a cosmology with $H_0=70$~km~s$^{-1}$~Mpc$^{-1}$, $\Omega_M=0.3$, and $\Omega_{\Lambda}=0.7$.
We adopt a Chabrier initial mass function (IMF; \citealt{chabrier03}).
Quoted uncertainties are at the $1\sigma$\ (68\%) confidence level.
The word ``gas'' in this work specifically means cold molecular gas (mostly H$_2$ and He), unless otherwise stated.

\section{Data and Analysis}
\label{sec:analyis}

\subsection{Targets and observations}
\label{sec:obs}
Our ALMA Cycle~7 program (2019.1.00678.S, PI: G.\ Yang) targeted seven bulge-dominated star-forming galaxies in the GOODS-South field, which has excellent multiwavelength coverage (see \S\ref{sec:sed}).
The targets were selected from the bulge-dominated sample in \cite{yang19}, and their basic properties are summarized in Table.~\ref{tab:basic}.
The bulge-dominated sample is classified using machine-learning morphological measurements \citep{huertas_company15} based on \textit{HST} $H$-band imaging \citep{grogin11, koekemoer11}.
The machine-learning approach is efficient and reliable and has been successfully applied to other fields beyond GOODS-South \citep[e.g.,][]{huertas_company15, ni21}.
\fst{This targeting approach is designed to select pure spheroidal galaxies that do not have a disk component (see, e.g., Fig.~2 of \citealt{yang19}),
and we discuss consequences of missing a disk component in \S\ref{sec:coev}.}
It is essential to focus on bulge-dominated galaxies, as the BHAR-SFR relation does not exist among other (e.g., disky and irregular) galaxies \citep{yang19}.
Consistently, in the local universe, BH masses are only tightly correlated with the masses of bulges rather than disks \citep[e.g.,][]{kormendy13}.

Our ALMA targets are selected to have secure optical spectroscopic-redshift (spec-$z$) measurements from the literature, and the redshifts span $z=0.5$--2.5 (see Table.~\ref{tab:basic}). 
The spec-$z$ measurements are necessary to locate the observed-frame CO frequencies and target them with ALMA.
The sample is also selected to have \spitzer/MIPS and/or \herschel\ $\fstm{(\rm S/N>3)}$ detections (\S\ref{sec:sed}), so that they are likely at the early star-forming phase after bulge formation (\S\ref{sec:intro}).
\fst{The \herschel\ flux uncertainties are based on Monte~Carlo simulations that account for both instrumental and confusion noise \citep[][]{elbaz11}.}
We are primarily interested in star-forming rather than quiescent bulges. 
This is because, for quiescent systems, both bulge and BH growth have essentially ceased according to the BHAR-bulge SFR relation \citep{yang19} and the system is already in place on the $\mbh$-$\mstar$ relation.\footnote{The program actually observed 10 targets in total. These 10 targets were originally selected using an earlier IR catalog, PEP \citep{lutz11}. However, the most recent catalog (\citealt{barro19}; \S\ref{sec:sed}), which carefully addresses source confusion, indicates 3 targets are actually quiescent, being undetected in the IR. Therefore, we only focus on the rest of the targets, i.e., 7 IR star-forming galaxies.} 
\fst{Five (out of seven) targets are classified as \xray\ AGNs in \cite{luo17}. 
This prevalence of AGNs is naturally expected from the BHAR-bulge SFR relation, as our targets are selected to be star-forming bulge-dominated galaxies.   
}

The ALMA data were taken on January 14 and 24, 2020, with the on-site perceptible water vapor (PWV) ranging from 4.6~mm to 6.6~mm.   
We target \coii\ and \coiii\ for $z<2$ and $z>2$ sources in our sample, respectively. 
These lines are within the frequency ranges covered ALMA bands 3 and 4. 
Among the four spectral windows (spw), one spw was placed centered on the line, and the other three covered continuum frequencies. 
These observations were performed on a 12-m array configuration of C-3 (maximum baseline $=0.5$~km), resulting in angular resolutions of 1.1--2.3~arcsec. 
The on-source exposure time was 24--28 minutes, reaching a $1\sigma$ continuum sensitivity of $\approx 0.02$--0.03~mJy~beam$^{-1}$. 

\begin{table*}
\begin{center}
\caption{Basic source properties}
\label{tab:basic}
\begin{tabular}{rrrrrrrrrrr} \hline\hline
ID & RA & DEC & $z$ & $z$ ref. & $\log \mstar$ & $\log \rm SFR$ & $\log L_{\rm X}$ & $\tgrow$ & $\tua$ \\
(1) & (2) & (3) & (4) & (5) & (6) & (7) & (8) & (9) & (10) \\
\hline
  528 & 53.113468 & $-27.933294$ & 1.089 &   \cite{cooper12} & $11.28\pm0.10$ & $ 1.58\pm0.04$ & $41.75\pm0.16$ & $ 5.09\pm1.25$ & 5.42  \\
 6278 & 53.060116 & $-27.852997$ & 1.540 &      \cite{suh15} & $10.97\pm0.05$ & $ 1.42\pm0.23$ & $43.00\pm0.07$ & $ 3.52\pm1.88$ & 4.11  \\
23845 & 53.097649 & $-27.715282$ & 2.142 &     \cite{coil15} & $10.96\pm0.02$ & $ 2.05\pm0.02$ & $43.41\pm0.09$ & $ 0.80\pm0.06$ & 3.02  \\
24210 & 53.071419 & $-27.717581$ & 0.566 &   \cite{cooper12} & $10.58\pm0.05$ & $ 1.89\pm0.02$ & $42.87\pm0.05$ & $ 0.49\pm0.06$ & 7.98  \\
24682 & 53.104096 & $-27.683758$ & 0.732 &   \cite{cooper12} & $10.78\pm0.02$ & $ 0.58\pm0.04$ & $42.98\pm0.05$ & $15.73\pm1.62$ & 6.99  \\
25573 & 53.139187 & $-27.694145$ & 1.044 & \cite{vanzella08} & $11.00\pm0.02$ & $ 1.14\pm0.04$ &             -- & $ 7.23\pm0.76$ & 5.58  \\
25998 & 53.137573 & $-27.700104$ & 2.453 &    \cite{barro13} & $11.05\pm0.07$ & $ 2.33\pm0.06$ & $43.68\pm0.08$ & $ 0.53\pm0.11$ & 2.63  \\
\hline
\end{tabular}
\end{center}
\begin{flushleft}
{\sc Note.} --- 
(1) Identification in the CANDELS catalog \citep{guo13}.
(2) \& (3) CANDELS J2000 coordinates.
(4) Optical spectroscopic redshift. 
(5) Redshift reference. 
(6) \& (7) Logarithmic stellar mass ($M_\odot$) and star formation rate ($M_\odot$~yr$^{-1}$) from our SED modelling (see \S\ref{sec:sed}).   
(8) Intrinsic \xray\ luminosity (erg~s$^{-1}$) based on the absorption-corrected \xray\ flux (see \S\ref{sec:sed}). ``--'' means \xray\ undetected. We note that $L_{\rm X}$ only reflects instantaneous AGN activity, not long-term average BH growth (see \S\ref{sec:intro}).
(9) The bulge stellar growth timescale (Gyr) as defined in Eq.~\ref{eq:tgrow}. 
\fst{(10) The Universe's age (Gyr) at the source's redshift.}
\end{flushleft}
\end{table*}

\subsection{Data reduction}
\label{sec:alma_data}
We requested and downloaded the calibrated Measurement Sets (MS) for our observations using the online Science Ready Data Products (SRDP) service.\footnote{\url{https://data.nrao.edu/portal/\#/}}
These MS data are produced by the ALMA Pipeline v6.1.2-7.
We use the Common Astronomy Software Applications (\casa) v6.2 package to further reduce these MS data. 

\fst{We employ the ``tclean'' function in \casa\ to produce the primary-beam (PB) corrected continuum images from the visibility data.
We include all available spw (masking potential line channels within $\pm 500$~km~s$^{-1}$ around the line center). 
We set the output image size to $20''\times 20''$ with a pixel scale of $0.2''$.
We then apply the ``imfit'' function (\casa) to these images to determine the source continuum position.
Five sources have converged fits but none of them is significant (S/N~$>3$), indicating the continuum emission is weak.
We present a quantitative analysis of the continuum fluxes in \S\ref{sec:cont}.}

\fst{For the line analysis, we first utilize tclean to transform the visibility data to an image cube for each source.}
We use the line spw and set the output spectral channel width to $50$~km~s$^{-1}$ \scd{(dataset native resolution $\approx 1$--3 km~s$^{-1}$)}, while the spaxel size and pixel scale are the same as above.
Using ``imcollapse'' (\casa), we then collapse the cube along the frequency axis including the line frequency (inferred from the optical redshift) $\pm 300$~km~s$^{-1}$.
Based on the resulting line image, we perform source detection with ``imfit'' (\casa), which searches for a Gaussian-shaped source. 
If a source is detected with signal-to-noise (S/N) $>3$, we adopt the imfit source position for spectrum-extraction below; otherwise, we adopt the CANDELS position (Table~\ref{tab:basic}).

\fst{The empirical choice of collapsing the data cube within $\pm 300$~km~s$^{-1}$ is to cover most of the line signal, as CO line width is typically $\lesssim 600$~km~s$^{-1}$ in the literature \citep[e.g.,][]{freundlich19, shangguan20}. 
Choosing an even larger width could dilute the S/N. 
We also test different ranges other than $\pm 300$~km~s$^{-1}$ (see \S\ref{sec:line_flux}) and find similar results. 
Admittedly, if a line center has a large shift greater than $300$~km~s$^{-1}$, our method could miss it.
However, from our extracted spectra, we do not find such strong shifts (see \S\ref{sec:line_flux}).
}

\begin{figure*}[h]
    \centering
	\includegraphics[width=0.368\columnwidth,valign=t]{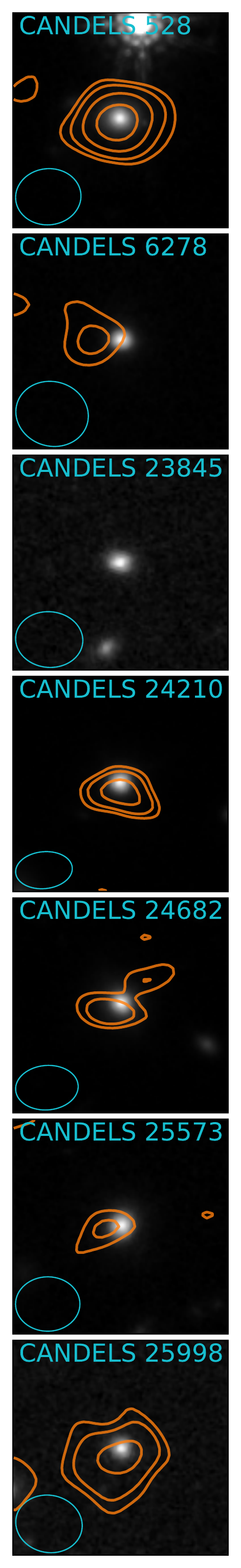}
	\includegraphics[width=1.635\columnwidth, height=0.76\paperheight,valign=t]{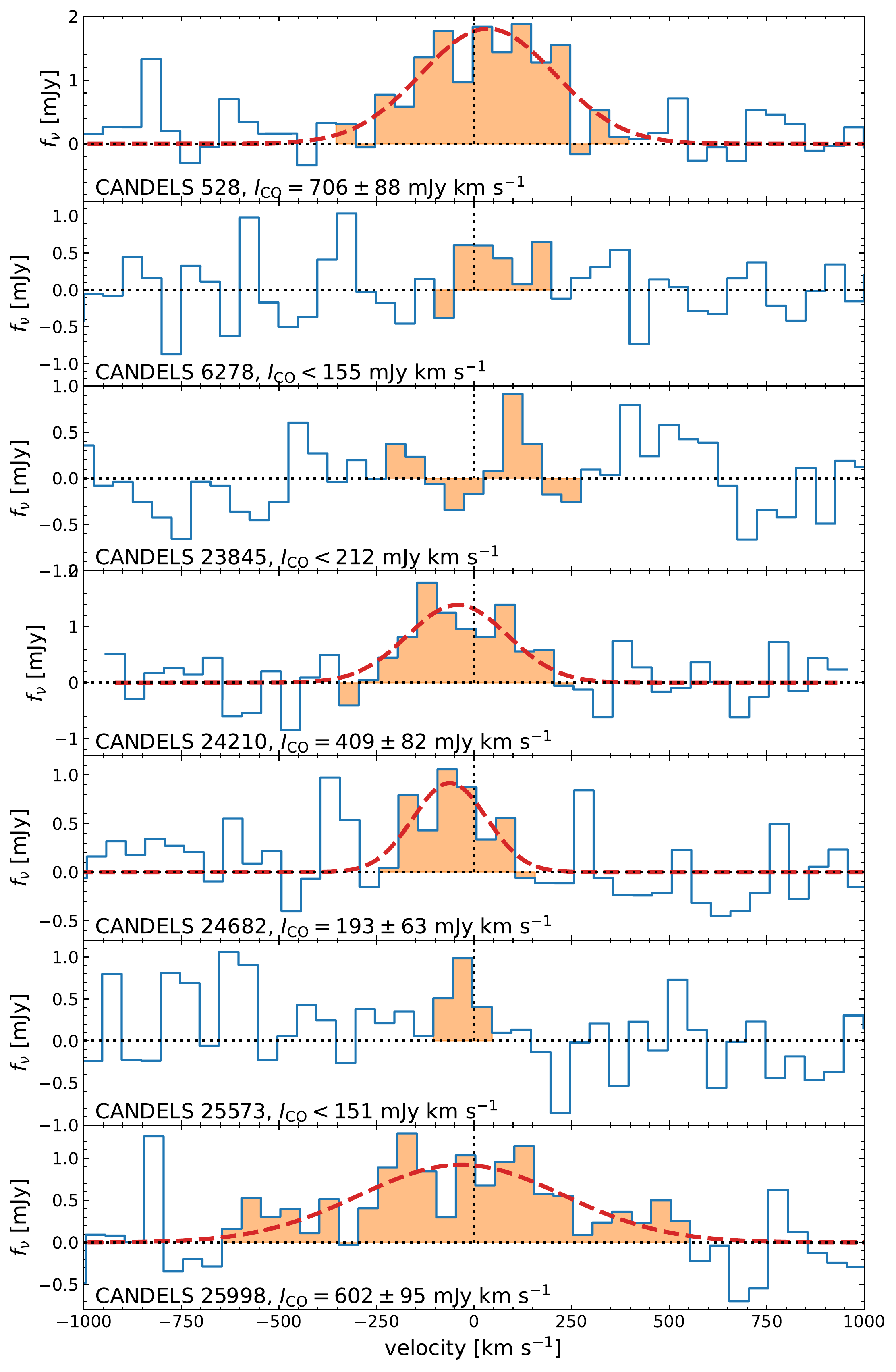}
    \caption{\textit{Left}: \textit{HST} $H$-band $7''\times 7''$ cutouts with contours of CO emission. 
    The contours are at the 2, 3, 5, and 8 sigma levels. 
    The beam profile is displayed at the bottom-left corner.
    There is no contour for CANDELS~23845 due to the weak signal of its CO-line map. 
    \textit{Right}: CO spectrum for each source. 
    The red dashed curve represents the best-fit Gaussian model (only displayed for CO detected sources). 
    The horizontal and vertical dotted lines indicate zero flux and velocity, respectively. 
    The orange shaded region indicates the integrated velocity range for the line-flux measurement.
    The measured line flux and its uncertainty (or upper limit) is labeled. 
    We consider the line as detected if S/N$>3$; we adopt the 3$\sigma$ uncertainties as upper limits for undetected sources (see Table~\ref{tab:alma}).
    }
    \label{fig:cutout}
\end{figure*}


\subsection{Line-flux measurements}
\label{sec:line_flux}
We extract the CO spectrum for each source with \casa's ``specflux'', utilizing a circular aperture with an area of $2\times$beam area.
\fst{Here, we choose a fixed aperture instead of an adaptive aperture based on, e.g., contours. 
This is because our targets are generally faint and affected by random noise significantly.
The latter adaptive approach, which favors ``positive'' noise and avoids ``negative'', may potentially introduce a positive bias.
This bias could lead to S/N over-estimation and even false detections.
}
\fst{The resulting spectra are displayed in Fig.~\ref{fig:cutout}.
From these spectra, there do not appear to be any significant line signals beyond 300~km~s$^{-1}$, supporting our source-detection method in \S\ref{sec:alma_data}.
}

We then fit the CO spectrum for each source with \casa's ``specfit'' employing a Gaussian model.
Here, we do not include a continuum component, because the continua are insignificant for all of our sources (see \S\ref{sec:cont}).
To avoid false detections, we require that the Gaussian center ($v_{\rm cent}$) is within $\pm 200$~km~s$^{-1}$, where the velocity zeropoint corresponds to the line frequency inferred from the optical redshift (and hereafter). 
The fits converge for 6 (out of 7) sources. 
We obtain their best-fit $v_{\rm cent}$ and full width half maximum (FWHM) from the specfit result. 
For each of the 6 sources, we estimate the velocity-integrated line flux (units: Jy~km~s$^{-1}$) by integrating the CO spectra over the range of ($v_{\rm cent}-$FWHM, $v_{\rm cent}+$FWHM).
For the other source (CANDELS~23845), we perform the integral over ($-300$~km~s$^{-1}$, $+300$~km~s$^{-1}$). 

To estimate the line-flux uncertainty for each source, we randomly place $\approx 20$ non-overlapping apertures around each source (avoiding its spectrum-extraction region). 
For each aperture, we then extract the spectrum and measure the velocity-integrated flux in the same way as above. 
We calculate the standard deviation of these fluxes. 
We repeat this process 100 times obtaining 100 standard-deviation values, and then calculate the median of these values.  
Finally, we adopt the median as the CO flux $1\sigma$ uncertainty for each source.   
If a source has CO-flux S/N$>3$, we consider the line as detected, and 4 out of 7 sources have CO detected. 
For the other 3 sources, we adopt 3$\times$ CO-flux errors as their upper limits. 
\fst{In \S\ref{sec:alma_data}, we collapse the data cubes over $\pm 300$~km~s$^{-1}$ for an initial source search.
We also test collapsing at other velocity ranges from $\pm 100$~km~s$^{-1}$ to $\pm 500$~km~s$^{-1}$ but do not find additional CO detections beyond the 4 sources.  
Therefore, we conclude that our results are not sensitive to the collapsed velocity range. 
}

Since the line flux above is measured within an aperture (``aperture flux'', hereafter), we need to do an aperture correction to account for the emission outside the aperture. 
To perform this task, we use a large circular aperture with an area of $4\times$beam area to estimate the ``total'' line flux for our highest S/N source (CANDELS~528 with $\rm S/N\approx 8$). 
We then divide this total flux by the aperture flux and adopt the result as our aperture-correction factor ($=1.12$). 
We multiply the line flux/error (or upper limit) by this aperture-correction factor.
\scd{We also test estimating the correction factor based on the other 3 detected sources. 
The resulting corrections are 1.10--1.28, similar to our adopted value (1.12). 
Our adopted correction is based on the highest S/N source, and thus should be the most reliable.
We note that our main results (\S\ref{sec:discuss}) are not sensitive to the choice of the aperture-correction factor, as all the values (from 1.10 to 1.28) are relatively small. 
}
Table~\ref{tab:alma} lists the final CO fluxes, errors, and upper limits for our sources. 
For the four CO-detected sources, we present spatial analyses of their CO emission in Appendix~\ref{sec:spat}. 
In brief, the angular resolutions of our ALMA observations are not sufficient to well resolve the detected CO emission.

From the measurements of CO line fluxes above, we estimate CO luminosities following \cite{solomon05}, i.e., 
\begin{equation}
\label{eq:lpco}
    \lpco(J,J-1) = 3.25\times 10^7 \ico \nu_{\rm obs}^{-2} D_L^2 (1+z)^{-3},
\end{equation}
where $\ico$ is the velocity-integrated line velocity in Jy~km~s$^{-1}$; $\nu_{\rm obs}$ is the line frequency in the observed frame; $D_L$ is the luminosity distance in Mpc; $\lpco(J,J-1)$ is in units of $\rm K\ km\ s^{-1}\ pc^2$.
Assuming $r_{21}=\lpco(2,1)/\lpco(1,0)=0.8$ and $r_{31}=\lpco(3,2)/\lpco(1,0)=0.5$ \citep[e.g.,][]{saintonge17, lamperti20}, we can convert the observed $\lpco(J,J-1)$ ($J=3$ or 2) to $\lpco(1,0)$ ($\lpco$ hereafter).
Finally, we estimate the molecular mass from 
\begin{equation}
\label{eq:aco}
    \mhh = \aco \lpco,
\end{equation}
where we assume the conversion factor $\aco=4.3\  M_\odot\ (\rm K\ km\ s^{-1}\ pc^2)^{-1}$ (we use these units for $\aco$, hereafter), a typical value adopted in the literature \citep[e.g.,][]{bolatto13, carilli13}. 
This adopted conversion factor includes the contribution from Helium and metals. 
We discuss the effects of this $\aco$ assumption in \S\ref{sec:coev}.
Table~\ref{tab:alma} lists the resulting $\lpco$ and $\mhh$.

\begin{table*}
\begin{center}
\caption{ALMA results}
\label{tab:alma}
\begin{tabular}{rrrrrrrrrrr} \hline\hline
ID & $J$ & Amp. & $v_{\rm cent}$ & $\Delta v$ & $\ico$ & $\log \mhh$ & $\tdep$ & $\mu$ & $S_\nu^{\rm cont}$ & $\log \mhh^{\rm cont}$ \\
(1) & (2) & (3) & (4) & (5) & (6) & (7) & (8) & (9) & (10) & (11) \\
\hline
  528 & 2 & $1.81\pm 0.22$ & $  35\pm  24$ & $ 404\pm  56$ & $ 706\pm  88$ & $10.78\pm0.05$ & $ 1.58\pm0.24$ & $ 0.31\pm0.08$ & $<  79$ & $<10.75$ \\
 6278 & 2 & --             & --            & --            & $< 155$       & $<10.41$       & $< 0.96$       & $< 0.27$        & $<  70$ & $<10.98$ \\
23845 & 3 & --             & --            & --            & $< 212$       & $<10.65$       & $< 0.40$       & $< 0.49$        & $<  80$ & $<10.95$ \\
24210 & 2 & $1.39\pm 0.24$ & $ -42\pm  26$ & $ 303\pm  62$ & $ 409\pm  82$ & $ 9.97\pm0.09$ & $ 0.12\pm0.02$ & $ 0.24\pm0.06$ & $<  93$ & $<10.25$ \\
24682 & 2 & $0.92\pm 0.24$ & $ -62\pm  28$ & $ 223\pm  67$ & $ 193\pm  63$ & $ 9.87\pm0.14$ & $ 1.93\pm0.65$ & $ 0.12\pm0.04$ & $<  95$ & $<10.49$ \\
25573 & 2 & --             & --            & --            & $< 151$       & $<10.07$       & $< 0.86$       & $< 0.12$        & $<  77$ & $<10.70$ \\
25998 & 3 & $0.92\pm 0.15$ & $ -33\pm  51$ & $ 618\pm 119$ & $ 602\pm  95$ & $11.21\pm0.07$ & $ 0.76\pm0.16$ & $ 1.43\pm0.32$ & $<  88$ & $<11.10$ \\
\hline
\end{tabular}
\end{center}
\begin{flushleft}
{\sc Note.} --- 
(1) Identification in the CANDELS catalog.
(2) Targeted CO transition ($J \rightarrow J -1$).
(3), (4), \& (5) Gaussian amplitude (mJy), central velocity (km~s$^{-1}$; relative to that from optical redshift), and FWHM (km~s$^{-1}$) from the fit of the CO spectrum. ``--'' indicates CO-undetected (S/N $<3$).
(6) Velocity-integrated CO flux or 3$\sigma$ upper limit (if S/N $<3$) in mJy~km~s$^{-1}$.
(7), (8) \& (9) Logarithmic gas mass ($M_\odot$) or 3$\sigma$ upper limit (inferred from $\lpco$; see \S\ref{sec:line_flux}), gas-depletion timescale (Gyr), and gas-to-stellar mass ratio.
(10) \& (11) 3$\sigma$ upper limit of continuum flux ($\mu$Jy) and corresponding logarithmic gas mass ($M_\odot$).
\end{flushleft}
\end{table*}

\subsection{Continuum emission}
\label{sec:cont}
\fst{In \S\ref{sec:alma_data}, we performed a source search on the ALMA continuum images but did not find any significant detections.}
Therefore, we adopt the CO line position (if available) or the CANDELS position for the continuum measurements.
As for the line-flux extraction, we also employ a circular aperture with area $=2\times$beam area to measure the continuum flux. 
We estimate the noise by randomly placing apertures and calculating the standard deviation of the resulting continuum fluxes (a similar procedure as in \S\ref{sec:line_flux}).
None of our sources has S/N$>3$ continuum fluxes, \fst{consistent with the result in \S\ref{sec:alma_data}.}
As in \S\ref{sec:line_flux}, we also adopt 3$\times$noise as the continuum upper limit for each source (listed in Table~\ref{tab:alma}). 
We then convert these flux upper limits to $\lef$ (rest-frame 850~$\mu$m luminosity) upper limits, assuming the Rayleigh-Jeans law ($L_\nu \propto \nu^{2}$) for $K$ corrections.
Finally, we constrain the continuum-based gas masses using the scaling relation, $\mhh^{\rm cont} = \lef / (\fstm{1.01 \times\ 10^{20}}\ \mathrm{erg\ s^{-1}\ Hz^{-1}}\ M_\odot^{-1})$, from \cite{scoville16, scoville17}.\footnote{\fst{The original scaling factor was $6.7 \times\ 10^{19}\ \mathrm{erg\ s^{-1}\ Hz^{-1}}\ M_\odot^{-1}$ assuming $\aco=6.5$ \citep{scoville16}. Here, we modify the scaling factor so that the relation becomes consistent with our assumed $\aco=4.3$ (\S\ref{sec:line_flux}).}}

The resulting $\mhh^{\rm cont}$ upper limits are listed in Table~\ref{tab:alma}.
\fst{For 3 out of the 4 CO-detected sources (except CANDELS~25998), the CO-based masses are consistent with the continuum-based constraints.} 
\fst{For CANDELS~25998, the CO-based mass is slightly higher than the continuum-based upper limit by $\approx 0.1$~dex (see \S\ref{sec:coev} for more discussion of this object).}
For the 3 CO-undetected sources, the CO-based mass constraints are tighter than continuum-based ones.
Therefore, our scientific discussions are primarily based on the CO measurements in \S\ref{sec:discuss}, unless otherwise stated. 

\subsection{Multiwavelength Spectral Energy Distributions}
\label{sec:sed}
We employ \cig\ v2022.0 \citep{roehlly14, boquien19, yang20, yang22} to perform SED modelling for our ALMA sources.
We compile multiwavelength broad-band photometric data from the Rainbow Cosmological Surveys Database.\footnote{\url{http://arcoirix.cab.inta-csic.es//Rainbow_navigator_public/}}
These data include 23 bands from $U$ to SPIRE~$500 \mu$m \citep{guo13, barro19}.
We also compile \xray\ photometry from the 7~Ms \textit{Chandra}\ Deep Field-South (\hbox{CDF-S}) catalog \citep{luo17}.
Six sources (all except CANDELS~25573) have \xray\ detections. 
Among them, five are detected in the hard band (2--7~keV) and we adopt their hard-band fluxes, which are less affected by obscuration compared to the full-band (0.5--7~keV) and soft-band (0.5--2~keV) fluxes.
Since \cig\ requires obscuration-corrected \xray\ fluxes as input \citep{yang20,yang22}, we apply obscuration corrections to these adopted fluxes. 
The corrections are estimated based on the absorption column densities ($N_H$) from \cite{luo17} and {\sc pimms}.\footnote{\url{https://cxc.harvard.edu/toolkit/pimms.jsp}}
One source (CANDELS~528) is only detected in the soft band, and we adopt the soft-band flux.\footnote{Obscuration correction is not possible for this source due to the single-band detection. 
However, we do not expect the obscuration is strong, because otherwise the detected band would likely be the hard band rather than the soft band.}

The model parameters in \cig\ are listed in Table~\ref{tab:cig}.
We adopt a delayed-$\tau$ model (\texttt{sfhdelayed} in \cig) for star formation history (SFH) and a \cite{bruzual03} model (\texttt{bc03} in \cig) for a simple stellar population (SSP).
\fst{We note that, when fitting the data, \cig\ automatically excludes the unphysical models in which the stellar age is older than the Universe's age.}
In \texttt{bc03}, we assume a \cite{chabrier03} IMF and a solar metallicity ($Z=0.02$).
We also include nebular emission (\texttt{nebular} in \cig) with default settings. 
We adopt the \cite{charlot00} model (\texttt{dustatt\_modified\_CF00} in \cig) for stellar attenuation. 
We allow the $V$-band attenuation varying from 0.2 to 3 mag, while leaving other parameters ($\mu_{V}$, $n_{\rm ISM}$, and $n_{\rm BC}$) at the default values.
We use the \cite{dale14} model (\texttt{dale2014} in \cig) for galactic dust emission. 
We allow three values for radiation-field slope, i.e., 1.5, 2, and 2.5.
We use the SKIRTOR model \citep[][\texttt{skirtor2016} in \cig]{stalevski12, stalevski16} for the AGN UV-to-IR emission.
We allow $\fracA$ (fractional AGN IR luminosity) to vary between 0 and 0.99. 
We set $\theta_{\rm AGN}$ (viewing angle) to 30$^\circ$ and 70$^\circ$, which are typical values for type~1 and type~2 AGNs, respectively \citep[e.g.,][]{yang20, ramos21}.
Finally, we include the \texttt{xray} module in \cig\ to account for AGN/galaxy \xray\ emission, and leave the related parameters at the default value(s).
The \cig\ configurations above lead to a total of 27,243,160 models (3,891,880 per source).
\fst{In addition to the photometry above, we also include the ALMA continuum upper limits (\S\ref{sec:cont}) in the fits.}

\begin{figure*}[ht]
    \centering
	\includegraphics[width=1.8\columnwidth]{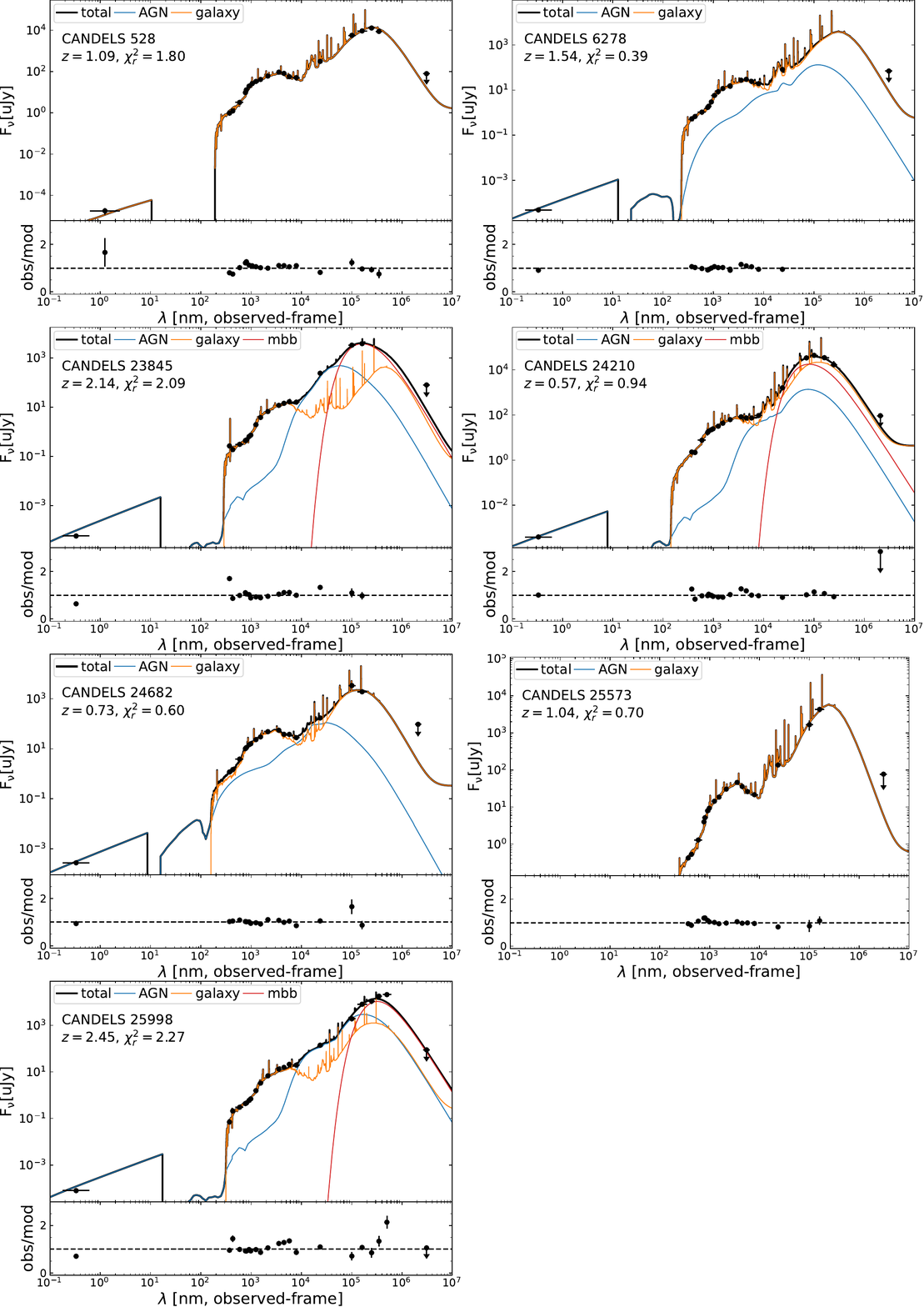}
    \caption{SED fits of our ALMA sources using \cig. 
    The black curve represents the best-fit SED model, while the blue, orange, and red curves indicate AGN, galaxy, and \texttt{mbb} (if present) components, respectively. 
    The redshift and reduced $\chi^2$ are labeled on each panel.
    }
    \label{fig:sed}
\end{figure*}

We run \cig\ with the configurations above. 
For 4 out of 7 sources, the fit quality is acceptable with reduced $\chi^2<2$ (see Fig.~\ref{fig:sed} for their best-fit SEDs).
We adopt the Bayesian output of $\mstar$ and SFR (Table.~\ref{tab:basic}).\footnote{The Bayesian output is a probability-weighted value that considers all models available. 
It is thus more robust than the best-fit output, which is based on a single model with minimum $\chi^2$.}
However, for CANDELS~23845, 24210, and 25998, the observed far-IR fluxes are systematically higher than the model values, leading to large reduced $\chi^2\approx 4$--5. 
This ``far-IR excess'' might be caused by a dust-enshrouded stellar population that is strongly attenuated at shorter wavelengths \citep[e.g.,][]{hodge16, buat19}. 
\fst{We note that CANDELS~23845, 24210, and 25998 are actually similar to the sample of \cite{buat19}, i.e., they are all massive ($\log(\mstar/M_\odot \gtrsim 10.5$) dust-rich ($\log(L_{\rm IR}/L_\odot) \gtrsim 12$) galaxies.
The \cite{buat19} sample's continuum emission is detected by ALMA, but ours is not (\S\ref{sec:cont}). 
We consider this difference is due to the different ALMA bands being used (band~6 vs.\ band~3) and that dust emission is stronger at shorter ALMA wavelengths due to the Rayleigh-Jeans law (e.g., Fig.~2 of \citealt{scoville16}).}

For these three sources, we include an additional \textit{ad hoc} modified (flexible emissivity) black body model (\texttt{mbb} in \cig) to account for this far-IR excess (Table~\ref{tab:cig}) and re-run \cig.
\scd{This \texttt{mbb} component is designed to model the far-IR emission from the dust-enshrouded stellar population as discussed above.
To account for possible temperature variance of this hidden population, we allow a cold model (50~K) and a hot model (100~K) in the \texttt{mbb} (see Table~\ref{tab:cig}).  
We allow the luminosity ratio, $L_{\rm mbb}$/$L_{\rm dale2014}$, to vary from 0 to 50. 
}
Indeed, the new fits have improved fit quality (reduced $\chi^2 \approx 1$--2; see Fig.~\ref{fig:sed}) compared to the previous fits. 
After adding the new component, the Akaike information criterion (AIC; \citealt{akaike74}) is reduced by $>10$, indicating the improvements with \texttt{mbb} are statistically significant \citep[e.g.,][]{burnham02}.

For the new fits, we do not adopt the SFRs directly from the \cig\ output, because the \cig\ SFRs do not account for the \texttt{mbb} contribution which dominates the galaxy IR luminosity ($\lgalir$).
We estimate SFRs based on the fitted total galaxy IR luminosity, \fst{excluding the AGN contribution} (\citealt{kennicutt98, salim07}), i.e., 
\begin{equation}
    {\rm SFR} = 1.09\times 10^{-10} \lgalir
\end{equation}
where SFR and $\lgalir$ are both in solar units.
We still adopt the output $\mstar$, assuming that the stellar mass is dominated by the main stellar population rather than the hidden population.
This assumption is justifiable, because the NIR data (rest-frame $\sim 1 \mu$m), a robust $\mstar$ indicator, can be fitted well with the main stellar population alone (see Fig.~\ref{fig:sed}).
We note that our main conclusions are unaffected, even if we miss some $\mstar$ contributed by the dust-enshrouded stellar population (see \S\ref{sec:coev}). 

We note that another approach to account for the far-IR excess is to adopt a flat dust-attenuation curve, which leads to significant attenuation at near-IR wavelengths \citep[e.g.,][]{buat19}.
This method effectively assumes that the far-IR excess comes from the strong attenuation in the near-IR. 
We also test this approach by allowing shallower ISM attenuation slopes ($n_{\rm ISM}$) ranging from $-0.3$ to $-0.7$ (disabling \texttt{mbb}). 
Indeed, the fit quality has been improved (reduced \hbox{$\chi^2 \approx 2$--2.5}) compared to the original fits (reduced $\chi^2 \approx 4$--5).
The resulting SFRs (also estimated from $\lgalir$) are similar to those from the \texttt{mbb} fits, with differences $\lesssim 0.1$~dex, but the $\mstar$ are systematically higher by $\approx 0.3$--0.6~dex.\footnote{\fst{Based on these alternative estimations of SFRs and $\mstar$, the sources are still classified as star-forming galaxies.}}
These effects are similar to what \cite{buat19} found. 
In \S\ref{sec:coev}, we discuss the effects to our main conclusions, if the flat-curve approach is adopted instead of the \texttt{mbb} one. 

The $\mstar$ and SFRs for all sources are listed in Table~\ref{tab:basic}.
Following \cite{ni21}, we classify a source as a star-forming galaxy if it satisfies $\rm SFR>SFR_{ MS}/10^{1.4}$, where $\rm SFR_{MS}$ is the main-sequence SFR as a function of $\mstar$ and redshift from \cite{whitaker12}. 
All of our 7 sources are classified as SF galaxies. 
\fst{From Table~\ref{tab:basic}, CANDELS~6278 has the largest SFR uncertainty (0.23~dex compared to $\lesssim 0.05$~dex for the other sources).\footnote{\fst{The error is calculated by \cig\ as the standard deviation of the marginalized SFR probability distribution, accounting for all available physical models \citep{boquien19}.}}
This is understandable, as CANDELS~6278 is the only object without \herschel\ detections in our sample (see Fig.~\ref{fig:sed}).
This level of SFR uncertainty (0.23~dex) without \herschel\ is realistic based on previous studies with \cig\ \citep[e.g.,][]{mountrichas21b}.
}

\begin{table*}
\centering
\caption{\cig\ model parameters}
\label{tab:cig}
\begin{tabular}{llll} \hline\hline
Module & Parameter & Symbol & Values \\
\hline
\multirow{2}{*}{\shortstack[l]{Star formation history\\
                               \texttt{sfhdelayed} }}
    & Stellar e-folding time & $\tau_{\rm star}$ & 0.5, 1, 2, 3, 4, 5 Gyr\\
    & Stellar age & $t_{\rm star}$  
            & \fst{0.2, 0.5,} 1, 2, 3, 4, 5 Gyr\\ 
\hline
\multirow{2}{*}{\shortstack[l]{Simple stellar population\\ 
    \texttt{bc03} }}
    & Initial mass function & $-$ & \cite{chabrier03} \\
    & Metallicity & $Z$ & 0.02 \\
\hline
\multirow{4}{*}{\shortstack[l]{Dust attenuation \\ 
                \texttt{dustatt\_modified\_CF00} }}
    & $V$-band attenuation in the ISM & $\avism$ & 0.2--1 (step 0.1), 1.5, 2, 2.5, 3 mag \\
    & $\avism/(\avism+\avbc)$ & $\mu_{V}$ & 0.44 \\
    & Slope of the ISM attenuation & $n_{\rm ISM}$ & $-0.7$ \\
    & Slope of the birth-cloud attenuation & $n_{\rm BC}$ & $-1.3$ \\
\hline
\multirow{4}{*}{\shortstack[l]{Galactic dust emission \\ \texttt{dale2014} \\ \texttt{mbb}$^{\rm a}$ }}
    & Slope in $dM_{\rm dust} \propto U^{-\alpha_{\rm dust}} dU$ & $\alpha_{\rm dust}$ & 1.5, 2, 2.5 \\
    & Ratio of $L_{\rm mbb}$ and $L_{\rm dale2014}$ & $\epsilon_{\rm mbb}$ & 0, 1, 2, 5, 10, 20, 50 \\
    & Temperature of mbb & $T_{\rm mbb}$ & 50, 100 K \\
    & Emissivity of mbb & $\beta$ & 1.5 \\
\hline
\multirow{3}{*}{\shortstack[l]{AGN (UV-to-IR) emission \\ \texttt{skirtor2016} }}
    & AGN contribution to IR luminosity & $\fracA$ & 0--0.9 (step 0.1), 0.99  \\
    & Viewing angle & $\theta_{\rm AGN}$ & 30$^\circ$, 70$^\circ$ \\
    & \multirow{1}{*}{\shortstack[l]{Polar-dust color excess}} & \multirow{1}{*}{\shortstack[l]{$E(B-V)_{\rm PD}$}} & \multirow{1}{*}{\shortstack[l]{0.03, 0.2, 0.4}} \\
\hline
    \multirow{2}{*}{\shortstack[l]{X-ray emission \\ \texttt{xray} }}
    & Maximum deviation from the $\ox$-$\luvr$ relation  & $|\Delta \ox|_{\rm max}$ & 0.2 \\
    & AGN \xray\ angle coefficients & $(a_1, a_2)$ &
        $(0.5, 0)$ \\
\hline
\end{tabular}
\begin{flushleft}
{\sc Note.} --- For parameters not listed here, we use the default values. 
(a) The \texttt{mbb} module is only used for CANDELS~23845, 24210, and 25998 to improve their fit quality (see \S\ref{sec:sed}).
\end{flushleft}
\end{table*}

\section{Discussion}
\label{sec:discuss}

\subsection{Implications for BH-bulge coevolution}
\label{sec:coev}
With the measurements of galaxy properties in \S\ref{sec:analyis}, we now discuss the implications for BH-bulge coevolution. 
First, we define a gas-depletion timescale as 
\begin{equation}
\label{eq:tdep}
\tdep \equiv \frac{\mhh}{\rm SFR},
\end{equation}
and bulge stellar growth timescale as 
\begin{equation}
\label{eq:tgrow}
\tgrow \equiv \frac{\mstar}{\rm SFR}.
\end{equation}
From this definition, $\tgrow$ represents the timescale needed to double the stellar mass in the future, given the current SFR.
\scd{Denoting the galaxy's redshift corresponds to a Universe's age of $t_0$ and assuming the SFR keeps constant with sufficient gas supply, then at \scd{cosmic time of} $t<t_0+\tgrow$, the stellar mass is predominantly assembled before $t_0$; at $t>t_0+\tgrow$, the mass is predominantly assembled after $t_0$.}
This stellar growth timescale is also a proxy for BH growth timescale due to
\begin{equation}
\label{eq:bh_grow}
 \tgrow = \frac{\mstar}{\rm SFR} = \frac{\mstar/300}{\rm SFR/300} \approx \frac{\mbh}{\rm BHAR} = \tgbh,
\end{equation}
where we apply the long-term average BHAR-SFR relation (BHAR$\approx \rm SFR/300$; see \S\ref{sec:intro}) and assume the local BH-bulge mass relation ($\mbh \approx \mstar /300$). 
We note that the system might not follow the local BH-bulge mass relation, and thus Eq.~\ref{eq:bh_grow} is only an approximation. 
If $\mbh$ is higher (lower) than the local BH-bulge mass relation, then $\tgbh$ should also be longer (shorter) than $\tgrow$.
\fst{The values of $\tgrow$ and $\tdep$ for our targets are listed in Tables~\ref{tab:basic} and \ref{tab:alma}, respectively.}

\begin{figure}[ht]
    \centering
	\includegraphics[width=\columnwidth]{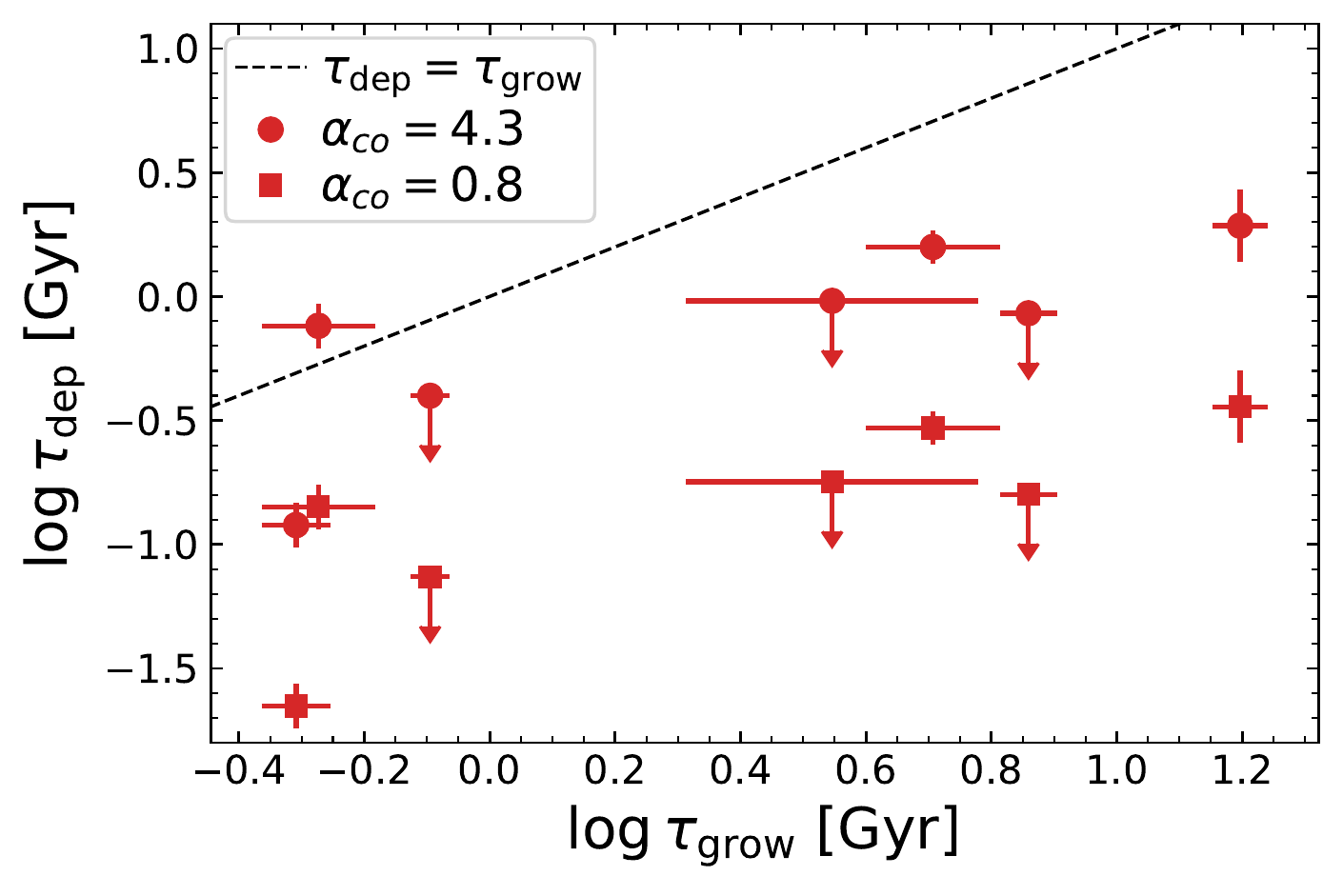}
    \caption{$\tdep$ versus $\tgrow$. 
    The red stars and squares are values estimated using $\aco=4.3$ (default) and $\aco=0.8$, respectively. 
    The downward-pointing arrows indicate 3$\sigma$ upper limits. 
    The black dashed lines indicate $\tdep=\tgrow$. 
    Except for CANDELS~25998 ($\aco=4.3$), other data points are all below the $\tdep=\tgrow$ line.
    }
    \label{fig:tvt}
\end{figure}

\fst{By comparing the two timescales of $\tdep$ and $\tgrow$, we can gain insight the BH-bulge coevolution.}
If $\tdep > \tgrow$, then the gas content can last sufficiently long for significant BH/bulge growth, \fst{as the BH/bulge mass predominantly forms during the bulge phase where the BHAR-SFR relation applies (Scenario 1 in Fig.~\ref{fig:scheme}).}
Otherwise, the gas is depleted quickly before significant BH/bulge growth \fst{and the system cannot significantly change its position on the $\mbh$-$\mstar$ diagram (Scenario 2 in Fig.~\ref{fig:scheme})}. 
Fig.~\ref{fig:tvt} compares $\tdep$ versus $\tgrow$ for our ALMA targets.
Six (out of seven) SF sources have $\tdep < \tgrow$, with the only exception being CANDELS~25998.
For these 6 sources, their $\tdep$ is more than 2 times shorter than $\tgrow$.

Our estimates of $\mhh$ assume a typical CO-to-gas conversion factor of $\aco=4.3$.
Some studies suggest a lower value of $\aco=0.8$ for compact galaxies \citep[e.g.,][]{bolatto13, carilli13}.
Considering that our bulge-dominated galaxies are generally compact \citep[e.g.,][]{ni19, ni21}, we also plot the result for the assumption of $\aco=0.8$ in Fig.~\ref{fig:tvt}. 
Under this alternative assumption, all seven sources, including CANDELS~25998, have $\tdep$ at least 4 times shorter than $\tgrow$.

\scd{$\tdep$ and $\tgrow$ are measures of what can happen in the future life of the galaxy assuming the current SFR.
But SFR is unlikely to be constant over cosmic time.}
From hydrodynamical simulations, SFR can be highly variable on relatively short timescales of $\sim 10$~Myr \citep[e.g.,][]{velazquez21}.
However, from Eqs.~\ref{eq:tdep} and \ref{eq:tgrow}, both $\tdep$ and $\tgrow$ are inversely proportional to SFR, and thus their ratio is independent of SFR, i.e.,
\begin{equation}
\label{eq:mu}
\frac{\tdep}{\tgrow} = \frac{\mhh}{\mstar} \equiv \mu,
\end{equation}
\scd{and $\mu$ is also known as the gas fraction of the galaxy.}
To compare $\tdep$ with $\tgrow$, it is equivalent to compare $\mu$ with unity. 
\fst{The advantage of $\mu$ is that it is not affected by the SFR variability mentioned above.
Aside from this mathematical feature, $\mu=\mhh/\mstar$ also has a useful physical meaning: it represents the amount of mass the BH/bulge can potentially grow in the future compared to the current mass. 
If $\mu$ is above unity, the ``potential'' mass will be dominant, supporting Scenario 1 in Fig.~\ref{fig:scheme}; otherwise, the ``current'' mass will be dominant, supporting Scenario 2 in Fig.~\ref{fig:scheme}.
}

\begin{figure}[ht]
    \centering
	\includegraphics[width=\columnwidth]{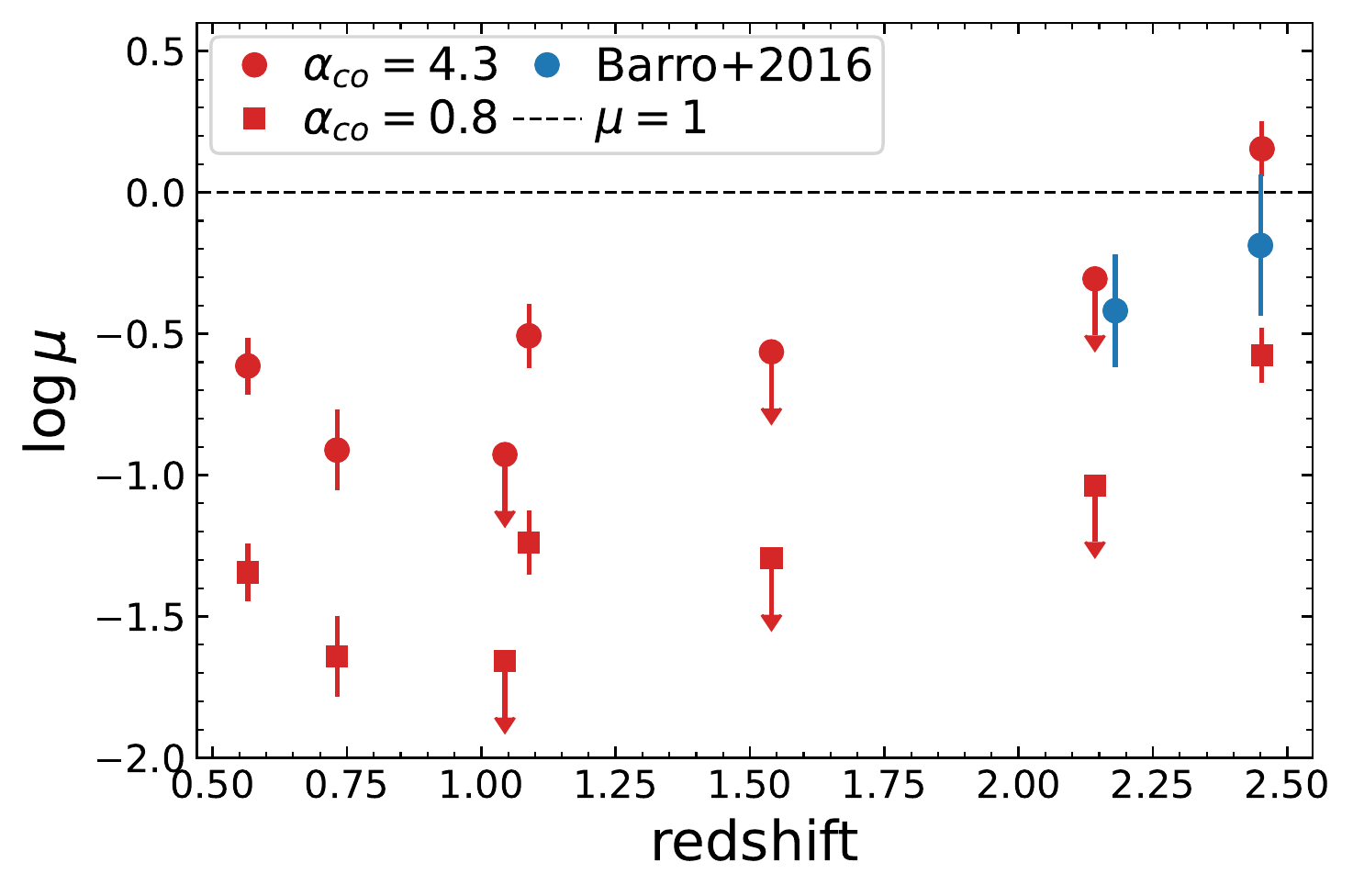}
    \caption{$\mu=\mhh/\mstar$ versus redshift. 
    The stars and squares are values estimated using $\aco=4.3$ (default) and $\aco=0.8$, respectively. 
    The downward-pointing arrows indicate 3$\sigma$ upper limits. 
    The dashed line indicates $\mu=1$.
    The two blue data points represent two bulge-dominated galaxies at $z=2$--2.5 from \cite{barro16}.
    The gas content of the bulge-dominated galaxies is generally low ($\mu < 1$ in most cases), insufficient to support significant bulge/BH growth.  
    }
    \label{fig:mu_vs_z}
\end{figure}

Fig.~\ref{fig:mu_vs_z} shows $\mu$ versus redshift for our ALMA targets, under the assumption of both $\aco=4.3$ and $\aco=0.8$. 
As expected, all sources have $\mu<1$ except CANDELS~25998 (for $\aco=4.3$).
Considering the upper-limit data points, the median values of $\mu$ are only $<0.26$ (for $\aco=4.3$) and $<0.048$ (for $\aco=0.8$), \fst{significantly smaller than unity under both $\aco$ assumptions.
Another source of systematic uncertainties is the assumed CO line ratios.
Our assumed values ($r_{21}=0.8$ and $r_{31}=0.5$) are typical among star-forming galaxies \citep[e.g.,][]{saintonge17, lamperti20}.
Some observations suggest that AGN hosts (the case for many of our targets; Fig.~\ref{fig:sed}) tend to have $r_{21}$ and $r_{31}$ higher than normal galaxies \citep[e.g.,][]{kirkpatrick19}. We note that adopting higher $r_{21}$ and $r_{31}$ would decrease the final gas masses and $\mu$, and thereby strengthen our main result above.}

\fst{In the discussion above, we assume that total $\mstar$ $\approx$ bulge $\mstar$, because our targets are bulge-dominated from \textit{HST} imaging (\S\ref{sec:obs}).
However, it is still possible that a minor disk component is missed, which could account for $<40\%$ of the total mass \citep[e.g.,][]{huertas_company16}. 
In the most extreme case (40\% mass from a missed disk component), our bulge $\mstar$ is overestimated by a factor of $1/(1-0.4) = 1.67$, and thus $\mu$ is underestimated by the same factor according to Eq.~\ref{eq:mu}.
Even this level of $\mu$ underestimation is not strong enough to overturn our main result, as most sources (except CANDELS~25998 assuming $\aco=4.3$) will still have $\mu<1$ after correcting for the underestimation (see Table.~\ref{tab:alma}).
}

In Fig.~\ref{fig:mu_vs_z}, we include two star-forming bulge-dominated sources also in the GOODS-South field from \cite{barro16}.\footnote{\cite{barro16} have six star-forming galaxies in total, but only two of them are bulge-dominated according to the morphological classifications of \cite{huertas_company15}.}
One source is CANDELS~21662 at $z=2.18$. 
The other is CANDELS~25998 at $z=2.45$, which is also in our sample. 
The \cite{barro16} $\mhh$ values were from $\mdust$ assuming a gas-to-dust ratio of 100, and their $\mdust$ estimation is based on SED modeling of ALMA continuum and other IR data.
\fst{For CANDELS~25998, the \cite{barro16} $\mhh$ ($10^{10.9}\ M_\odot$) is consistent with our continuum-based $\mhh$ constraint ($<10^{11.1}\ M_\odot$), although the two measurements are based on different frequencies and assumptions.}
The two $\mu$ values from \cite{barro16} are both below unity, consistent with our sample. 
For CANDELS~25998, the \cite{barro16} $\mstar$ estimation is similar to ours ($10^{11.07}$ vs.\ $10^{11.05}\ M_\odot$), and thus the cause of the $\mu$ difference between \cite{barro16} and ours (the rightmost points in Fig.~\ref{fig:mu_vs_z}) is mostly related to the $\mhh$ measurements. 
The \cite{barro16} $\mu$ is between our values based on $\aco=0.8$ and $\aco=4.3$, suggesting that the intrinsic $\aco$ for this source is within 0.8--4.3.
But we note that the uncertainties of \cite{barro16} $\mu$ are relatively large, and both of our $\mu$ values are consistent with their measurements at a $2\sigma$ confidence level. 

Our results above suggest that, without gas replenishment (see \S\ref{sec:repl}), the cold molecular gas of the SF bulge-dominated galaxies will be depleted before significant BH/bulge growth. 
Their $\mbh/\mstar$ ratios remain largely unchanged during the bulge-evolution phase until $z=0$ (see \S\ref{sec:intro}). 
Therefore, it is likely that the $\mbh$-$\mstar$ relation has already formed at the beginning of the bulge phase and they maintain this relation until $z=0$ (i.e., Scenario~2 in Fig.~\ref{fig:scheme}). 
The detailed formation mechanisms are unknown, subject to investigations from both theoretical and observational approaches \citep[e.g.,][]{huertas_company18, hopkins22}.
The BHAR-bulge SFR relation has the role of maintaining the BH-bulge mass correlation, but not creating it. 
We caution that our sample is limited to $z=0.5$--2.5. 
It is possible that this conclusion might change at higher redshift, considering that cold gas tends to be more abundant toward the early universe. 
Also, from Fig.~\ref{fig:mu_vs_z}, the $z>2$ sources tend to have higher $\mu$ than sources at lower redshifts, although our sample size not sufficiently large to robustly test this trend. 

In our SED fits (\S\ref{sec:sed}), we employ an \textit{ad hoc} \texttt{mbb} component to account for the FIR excess in three sources. 
We might underestimate $\mstar$ as we do not account for the mass contribution from the hidden stellar population. 
Therefore, the actual $\mu$ values might be even lower than our case, further strengthen our main conclusion.
The FIR excess can also be addressed by adopting a flat attenuation curve. 
We note that this approach would also lead to higher $\mstar$ estimations (see \S\ref{sec:sed}) and thereby lower $\mu$.
In summary, we consider that our main conclusion is not affected qualitatively by the details of the SED-fitting procedure. 

\subsection{Gas replenishment}
\label{sec:repl}
The discussion in \S\ref{sec:coev} assumes that the gas content in our bulge-dominated galaxies is primarily consumed by star formation without replenishment. 
However, if gas replenishment is common and supplies $\mhh$ by a typical factor of $\gtrsim 4$ (as median $\mu<0.26$ for $\aco=4.3$ in \S\ref{sec:coev}), then the BH/bulge growth can sustain for a longer timescale than our estimated $\tdep$.

\begin{figure}[ht]
    \centering
	\includegraphics[width=\columnwidth]{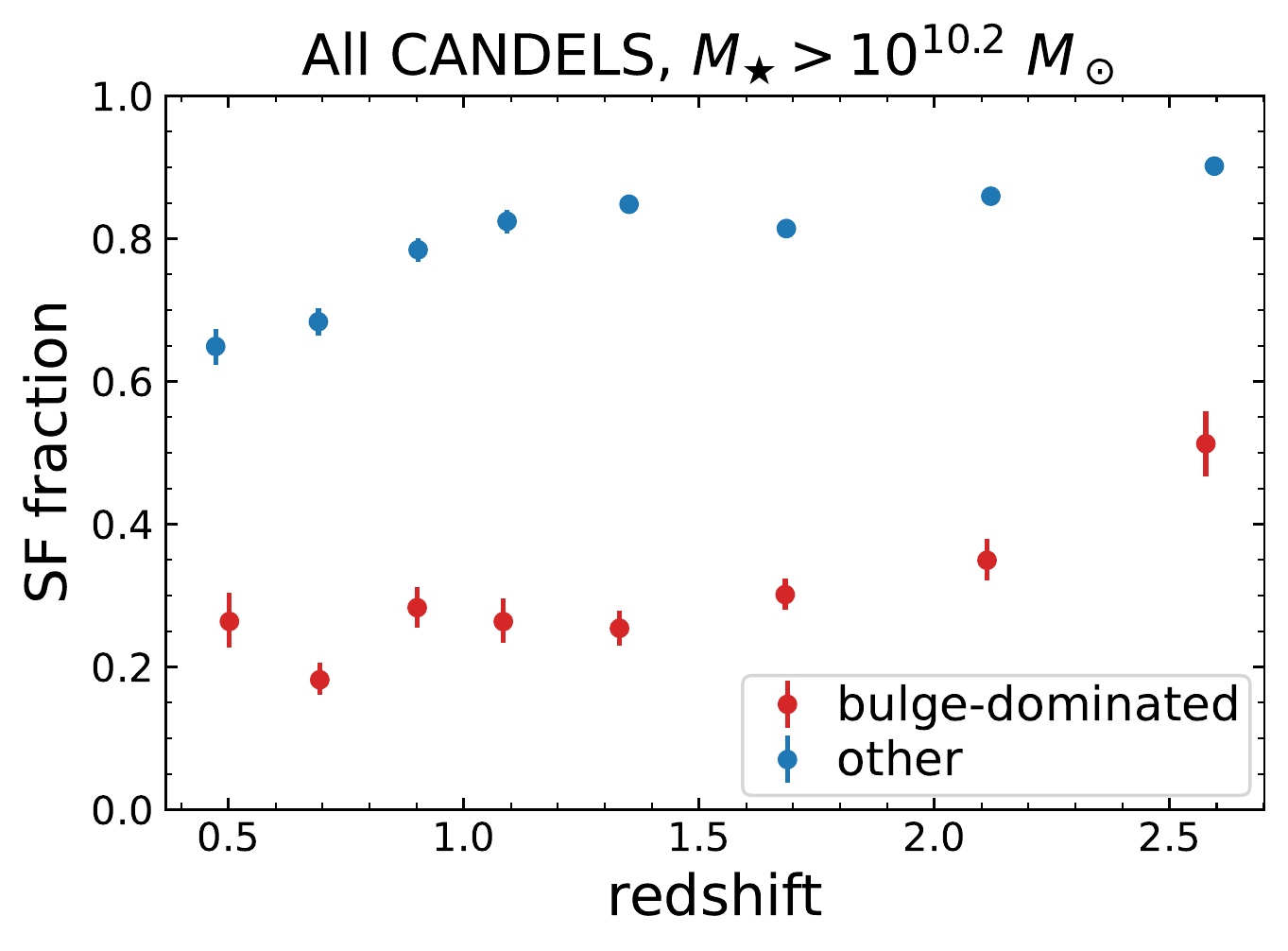}
    \caption{Fraction of star-forming galaxies ($\mstar > 10^{10.2}\ M_\odot$) as a function of redshift. 
    The red data points represent bulge-dominated galaxies, and the blue ones are for the other galaxies that are not bulge-dominated. 
    The error bars represent 1$\sigma$ binomial uncertainties calculated with the ``binom\_conf\_interval'' function of {\sc astropy}.
    The star-forming fraction is low (\hbox{$\approx 0.2$--0.3} at \hbox{$z\approx 0.5$--2}) for bulge-dominated galaxies, indicating that gas replenishment is not common for them.  
    }
    \label{fig:sf_frac_vs_z}
\end{figure}

This possibility can be qualitatively investigated by studying the fraction of star-forming galaxies, as common gas replenishment means widespread SF activity.    
Fig.~\ref{fig:sf_frac_vs_z} displays SF fraction versus redshift among bulge-dominated galaxies in all five CANDELS fields. 
The sample's properties (redshift, $\mstar$, and SFR) are compiled/estimated by \cite{yang19}. 
Here, we classify SF versus quiescent using the method in \S\ref{sec:sed}.
The SF fraction values are calculated for galaxies more massive than $10^{10.2}\ M_\odot$, above which the bulge-dominated sample is complete up to $z\approx 3$ \citep{yang19}.

From Fig.~\ref{fig:sf_frac_vs_z}, the SF fraction for bulge-dominated galaxies is generally low (e.g., $\approx 0.2$--0.3 at $z\approx 1$--2).
The low SF fraction suggests that bulge-dominated galaxies do not have prevalent strong gas replenishment, which is required to maintain their SF activity.
In comparison, we also plot the SF fraction for galaxies that are not bulge-dominated (e.g., disky or irregular) in Fig.~\ref{fig:sf_frac_vs_z}.
These galaxies tend to have high SF fractions (e.g., $\approx 0.8$--0.9 at $z\approx 1$--2), indicating prevalent gas replenishment among them.

We caution that the argument above only qualitatively suggests that bulge-dominated galaxies have weaker gas replenishment than non-bulge-dominated galaxies.
It does not rule out intermittent gas accretion among bulge-dominated galaxies.
Intermittent gas accretion might lead to sporadic star formation, also consistent with the low but non-zero SF fraction among bulge-dominated galaxies (see Fig.~\ref{fig:sf_frac_vs_z}). 
A quantitative assessment of intermittent gas accretion is beyond the scope of this work.

\section{Summary and Future Prospects}
\label{sec:sum}
In this work, we have presented the ALMA observations of 7 bulge-dominated star-forming galaxies at $z=0.5$--2.5.
Our main results are summarized below. 

\begin{itemize}
    \item We have reduced the ALMA data and measured the \coii\ or \coiii\ fluxes (see \S\ref{sec:analyis}).
    We have detected the CO lines from 4 sources at $>3\sigma$ levels, and we have estimated 3$\sigma$ upper limits for the other sources (\S\ref{sec:line_flux}). 
    From these results, we have inferred molecular gas masses (or upper limits) assuming $\aco=4.3$. 
    By fitting the existing multiwavelength data with \cig, we have estimated the stellar masses and star formation rates for our ALMA targets (\S\ref{sec:sed}).
    
    \item The gas-depletion timescales are at least 2 times shorter than the corresponding bulge/BH growth timescales for most sources, except for CANDELS~25998 (see \S\ref{sec:coev}). 
    The median value of $\mu$ is only $<0.26$. 
    If we assume $\aco=0.8$ (a typical value for compact galaxies) instead of $\aco=4.3$,
    all sources have $\mu<0.25$ (i.e., $\tdep$ is $>4$ times shorter than $\tgrow$) with a median $<0.048$.
    The ALMA continuum measurement from \cite{barro16} also suggests $\mu<1$ for two bulge-dominated sources (including CANDELS~25998).
    Therefore, we conclude that, without strong gas replenishment (supplying $\mhh$ by a factor of $\gtrsim 4$), the observed gas content of the SF bulges is generally insufficient to support significant bulge/BH growth. 
    
    \item To assess gas replenishment, we have estimated the SF fraction for a mass-complete sample of CANDELS galaxies (see \S\ref{sec:repl}). 
    The SF fraction for bulge-dominated galaxies is much lower than that for non-bulge-dominated galaxies (e.g., \hbox{$\approx 0.2$--0.3} versus \hbox{$\approx 0.8$--0.9} at $z=1$--2).
    The low SF fraction of bulge-dominated galaxies indicates that gas replenishment is not a common process among them. 
    We caution that our qualitative argument cannot rule out weak intermittent gas accretion among bulge-dominated galaxies.
    
    \item Our overall results indicate that the $\mbh$-$\mstar$ relation has already formed at the beginning of the bulge evolution phase (Scenario 2 in Fig.~\ref{fig:scheme}). 
    The systems then maintain this relation until $z=0$.
    In other words, the BHAR-bulge SFR relation has the role of maintaining the BH-bulge mass correlation, but not creating it. 
    Therefore, it will be useful to study the BH-galaxy coevolution in the pre-bulge phase, which might reveal the mysterious origin of the $\mbh$-$\mstar$ relation. 
    Such a study requires reliable techniques, probably machine learning trained by hydrodynamical simulations \citep[e.g.,][]{huertas_company18}, to select pre-bulge samples from high-resolution images. 

\end{itemize}

Finally, we note that our sample size is limited (7 sources), and they are all below $z=2.5$.
\fst{If this small sample is somehow biased toward the late stage of star formation  (\S\ref{sec:alma_data}), then our main conclusion could be altered.
Future (sub)mm observations of a much larger ($\gtrsim 100$) bulge-dominated sample over a wider parameter space (especially at $z>2.5$) will naturally address this potential issue and further test our conclusion.}
High-redshift morphological classifications, which are necessary to select bulge-dominated sources, will be available in the near future with the advance of \textit{JWST}\ extragalactic surveys. 
ALMA or other (sub)mm facilities can perform follow-up observations of the \textit{JWST}-selected targets. 

\section*{Acknowledgments}
We thank the referee for helpful feedback that improved this work.
We thank Guillermo Barro and Ian Smail for helpful discussions. 
GY, CP, JSS, JLW, and CPZ acknowledge support from the George P.\ and Cynthia Woods Mitchell Institute for Fundamental Physics and Astronomy at Texas A\&M University.
WNB acknowledges support from Chandra X-ray Center grant GO9-20099X and the V.M. Willaman Endowment.
DMA thanks the Science and Technology Facilities Council for support through grant code ST/T000244/1.
MB gratefully acknowledges support by the ANID BASAL project FB210003 and from the FONDECYT regular grant 1211000.
The authors acknowledge the Texas A\&M University Brazos HPC cluster and Texas A\&M High Performance Research Computing Resources (HPRC, http://hprc.tamu.edu) that contributed to the research reported here.
We thank the ALMA helpdesk for their help with ALMA data retrieval and reduction.  
This paper makes use of the following ALMA data: ADS/JAO.ALMA\#2019.1.00678.S. 
The Joint ALMA Observatory is operated by ESO, AUI/NRAO and NAOJ.
The National Radio Astronomy Observatory is a facility of the National Science Foundation operated under cooperative agreement by Associated Universities, Inc.
This work has made use of the Rainbow Cosmological Surveys Database, which is operated by the Centro de Astrobiolog\'ia (CAB/INTA), partnered with the University of California Observatories at Santa Cruz (UCO/Lick, UCSC).

\software{
{\sc astropy} \citep[v4.2;][]{astropy},
\cig\ \citep[v2022.0;][]{boquien19, yang20, yang22},
{\sc casa} \citep[v6.2.0;][]{mcmullin07}.
}

\appendix
\section{Spatial analysis}
\label{sec:spat}
\begin{figure*}[ht]
    \centering
	\includegraphics[width=2\columnwidth]{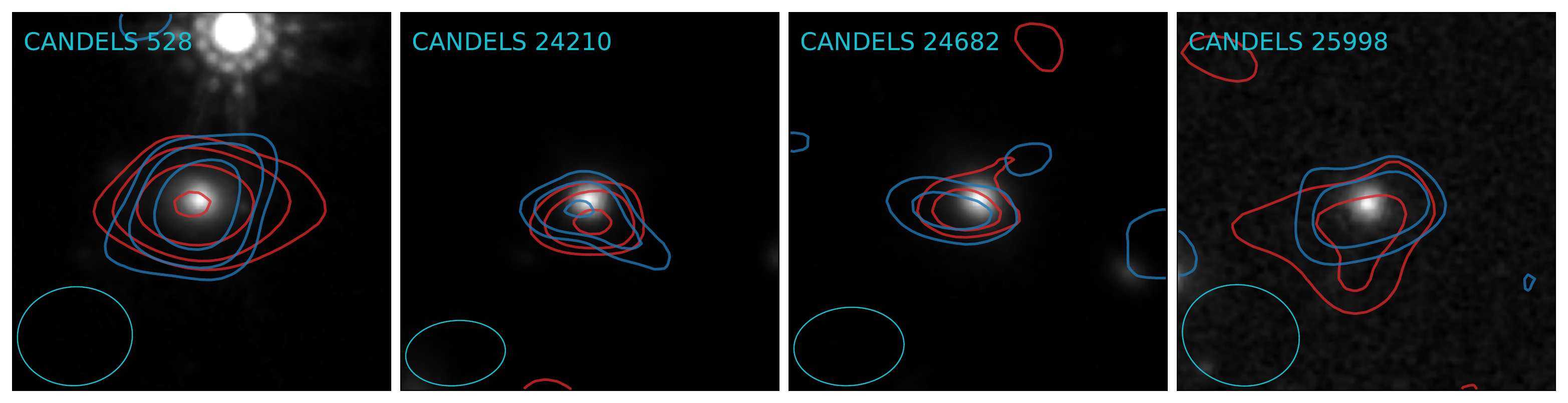}
    \caption{\textit{HST} $H$-band $7''\times 7''$ cutouts with contours of CO emission for the four CO-detected sources. 
    The blue and red contours are from the blue and red halves of the line, respectively.
    The contours are at the 2, 3, 5, and 8 sigma levels. 
    The beam profile is displayed at the bottom-left corner.
    }
    \label{fig:cutout_blue_red}
\end{figure*}

For the 4 CO-detected sources (\S\ref{sec:line_flux}), we further perform spatial analysis for their lines. 
For each of the 4 sources, we divide the line into blue and red halves and make a line image for each half. 
The resulting line-image contours are displayed in Fig.~\ref{fig:cutout_blue_red}. 
From this figure, only CANDELS~24210 appears to have (slightly) separated blue versus red contours. 
The angular distance between the blue and red peaks is $\approx 0.5''$.
On the other hand, the positional uncertainty for the blue/red line emission is $\approx 0.6 \theta \rm (S/N)^{-1} \approx 0.26 ''$, where $\theta \approx 1.5''$ is the synthesized beam FWHM \citep[e.g.,][]{ivison07}.
Therefore, the $\approx 0.5''$ separation is marginally significant at a $\approx 2\sigma$ level. 

The results above indicate that our current ALMA data are not able to spatially resolve the line-emitting regions in general. 
This is understandable considering the relatively large beam sizes compared to the $H$-band profiles (see Fig.~\ref{fig:cutout_blue_red}).
High-resolution ALMA runs are necessary to probe the spatial distribution of the CO emission. 

\bibliography{all}{}

\begin{thebibliography}{}
\expandafter\ifx\csname natexlab\endcsname\relax\def\natexlab#1{#1}\fi
\providecommand{\url}[1]{\href{#1}{#1}}
\providecommand{\dodoi}[1]{doi:~\href{http://doi.org/#1}{\nolinkurl{#1}}}
\providecommand{\doeprint}[1]{\href{http://ascl.net/#1}{\nolinkurl{http://ascl.net/#1}}}
\providecommand{\doarXiv}[1]{\href{https://arxiv.org/abs/#1}{\nolinkurl{https://arxiv.org/abs/#1}}}

\bibitem[{{Aird} {et~al.}(2018){Aird}, {Coil}, \& {Georgakakis}}]{aird18}
{Aird}, J., {Coil}, A.~L., \& {Georgakakis}, A. 2018, \mnras, 474, 1225,
  \dodoi{10.1093/mnras/stx2700}

\bibitem[{Akaike(1974)}]{akaike74}
Akaike, H. 1974, IEEE transactions on automatic control, 19, 716

\bibitem[{{Astropy Collaboration} {et~al.}(2018){Astropy Collaboration},
  {Price-Whelan}, {Sip{\H{o}}cz}, {G{\"u}nther}, {Lim}, {Crawford}, {Conseil},
  {Shupe}, {Craig}, \& {Dencheva}}]{astropy}
{Astropy Collaboration}, {Price-Whelan}, A.~M., {Sip{\H{o}}cz}, B.~M., {et~al.}
  2018, \aj, 156, 123, \dodoi{10.3847/1538-3881/aabc4f}

\bibitem[{{Barro} {et~al.}(2013){Barro}, {Faber}, {P{\'e}rez-Gonz{\'a}lez},
  {Koo}, {Williams}, {Kocevski}, {Trump}, {Mozena}, {McGrath}, {van der Wel},
  {Wuyts}, {Bell}, {Croton}, {Ceverino}, {Dekel}, {Ashby}, {Cheung},
  {Ferguson}, {Fontana}, {Fang}, {Giavalisco}, {Grogin}, {Guo}, {Hathi},
  {Hopkins}, {Huang}, {Koekemoer}, {Kartaltepe}, {Lee}, {Newman}, {Porter},
  {Primack}, {Ryan}, {Rosario}, {Somerville}, {Salvato}, \& {Hsu}}]{barro13}
{Barro}, G., {Faber}, S.~M., {P{\'e}rez-Gonz{\'a}lez}, P.~G., {et~al.} 2013,
  \apj, 765, 104, \dodoi{10.1088/0004-637X/765/2/104}

\bibitem[{{Barro} {et~al.}(2016){Barro}, {Kriek}, {P{\'e}rez-Gonz{\'a}lez},
  {Trump}, {Koo}, {Faber}, {Dekel}, {Primack}, {Guo}, {Kocevski},
  {Mu{\~n}oz-Mateos}, {Rujopakarn}, \& {Seth}}]{barro16}
{Barro}, G., {Kriek}, M., {P{\'e}rez-Gonz{\'a}lez}, P.~G., {et~al.} 2016,
  \apjl, 827, L32, \dodoi{10.3847/2041-8205/827/2/L32}

\bibitem[{{Barro} {et~al.}(2019){Barro}, {P{\'e}rez-Gonz{\'a}lez}, {Cava},
  {Brammer}, {Pandya}, {Eliche Moral}, {Esquej}, {Dom{\'\i}nguez-S{\'a}nchez},
  {Alcalde Pampliega}, {Guo}, {Koekemoer}, {Trump}, {Ashby}, {Cardiel},
  {Castellano}, {Conselice}, {Dickinson}, {Dolch}, {Donley}, {Espino Briones},
  {Faber}, {Fazio}, {Ferguson}, {Finkelstein}, {Fontana}, {Galametz},
  {Gardner}, {Gawiser}, {Giavalisco}, {Grazian}, {Grogin}, {Hathi}, {Hemmati},
  {Hern{\'a}n-Caballero}, {Kocevski}, {Koo}, {Kodra}, {Lee}, {Lin}, {Lucas},
  {Mobasher}, {McGrath}, {Nandra}, {Nayyeri}, {Newman}, {Pforr}, {Peth},
  {Rafelski}, {Rodr{\'\i}guez-Munoz}, {Salvato}, {Stefanon}, {van der Wel},
  {Willner}, {Wiklind}, \& {Wuyts}}]{barro19}
{Barro}, G., {P{\'e}rez-Gonz{\'a}lez}, P.~G., {Cava}, A., {et~al.} 2019, \apjs,
  243, 22, \dodoi{10.3847/1538-4365/ab23f2}

\bibitem[{{Bolatto} {et~al.}(2013){Bolatto}, {Wolfire}, \& {Leroy}}]{bolatto13}
{Bolatto}, A.~D., {Wolfire}, M., \& {Leroy}, A.~K. 2013, \araa, 51, 207,
  \dodoi{10.1146/annurev-astro-082812-140944}

\bibitem[{{Boquien} {et~al.}(2019){Boquien}, {Burgarella}, {Roehlly}, {Buat},
  {Ciesla}, {Corre}, {Inoue}, \& {Salas}}]{boquien19}
{Boquien}, M., {Burgarella}, D., {Roehlly}, Y., {et~al.} 2019, \aap, 622, A103,
  \dodoi{10.1051/0004-6361/201834156}

\bibitem[{{Brandt} \& {Yang}(2021)}]{brandt21}
{Brandt}, W.~N., \& {Yang}, G. 2021, arXiv e-prints, arXiv:2111.01156.
\newblock \doarXiv{2111.01156}

\bibitem[{{Bruzual} \& {Charlot}(2003)}]{bruzual03}
{Bruzual}, G., \& {Charlot}, S. 2003, \mnras, 344, 1000,
  \dodoi{10.1046/j.1365-8711.2003.06897.x}

\bibitem[{{Buat} {et~al.}(2019){Buat}, {Ciesla}, {Boquien}, {Ma{\l}ek}, \&
  {Burgarella}}]{buat19}
{Buat}, V., {Ciesla}, L., {Boquien}, M., {Ma{\l}ek}, K., \& {Burgarella}, D.
  2019, \aap, 632, A79, \dodoi{10.1051/0004-6361/201936643}

\bibitem[{Burnham \& Anderson(2002)}]{burnham02}
Burnham, K., \& Anderson, D. 2002, Model selection and multimodel inference: a
  practical information-theoretic approach, 2, 49

\bibitem[{{Carilli} \& {Walter}(2013)}]{carilli13}
{Carilli}, C.~L., \& {Walter}, F. 2013, \araa, 51, 105,
  \dodoi{10.1146/annurev-astro-082812-140953}

\bibitem[{{Chabrier}(2003)}]{chabrier03}
{Chabrier}, G. 2003, \apjl, 586, L133, \dodoi{10.1086/374879}

\bibitem[{{Charlot} \& {Fall}(2000)}]{charlot00}
{Charlot}, S., \& {Fall}, S.~M. 2000, \apj, 539, 718, \dodoi{10.1086/309250}

\bibitem[{{Coil} {et~al.}(2015){Coil}, {Aird}, {Reddy}, {Shapley}, {Kriek},
  {Siana}, {Mobasher}, {Freeman}, {Price}, \& {Shivaei}}]{coil15}
{Coil}, A.~L., {Aird}, J., {Reddy}, N., {et~al.} 2015, \apj, 801, 35,
  \dodoi{10.1088/0004-637X/801/1/35}

\bibitem[{{Conselice}(2014)}]{conselice14}
{Conselice}, C.~J. 2014, \araa, 52, 291,
  \dodoi{10.1146/annurev-astro-081913-040037}

\bibitem[{{Cooper} {et~al.}(2012){Cooper}, {Yan}, {Dickinson}, {Juneau},
  {Lotz}, {Newman}, {Papovich}, {Salim}, {Walth}, {Weiner}, \&
  {Willmer}}]{cooper12}
{Cooper}, M.~C., {Yan}, R., {Dickinson}, M., {et~al.} 2012, \mnras, 425, 2116,
  \dodoi{10.1111/j.1365-2966.2012.21524.x}

\bibitem[{{Dale} {et~al.}(2014){Dale}, {Helou}, {Magdis}, {Armus},
  {D{\'\i}az-Santos}, \& {Shi}}]{dale14}
{Dale}, D.~A., {Helou}, G., {Magdis}, G.~E., {et~al.} 2014, \apj, 784, 83,
  \dodoi{10.1088/0004-637X/784/1/83}

\bibitem[{{Elbaz} {et~al.}(2011){Elbaz}, {Dickinson}, {Hwang},
  {D{\'{\i}}az-Santos}, {Magdis}, {Magnelli}, {Le Borgne}, {Galliano},
  {Pannella}, {Chanial}, {Armus}, {Charmandaris}, {Daddi}, {Aussel}, {Popesso},
  {Kartaltepe}, {Altieri}, {Valtchanov}, {Coia}, {Dannerbauer}, {Dasyra},
  {Leiton}, {Mazzarella}, {Alexander}, {Buat}, {Burgarella}, {Chary}, {Gilli},
  {Ivison}, {Juneau}, {Le Floc'h}, {Lutz}, {Morrison}, {Mullaney}, {Murphy},
  {Pope}, {Scott}, {Brodwin}, {Calzetti}, {Cesarsky}, {Charlot}, {Dole},
  {Eisenhardt}, {Ferguson}, {F{\"o}rster Schreiber}, {Frayer}, {Giavalisco},
  {Huynh}, {Koekemoer}, {Papovich}, {Reddy}, {Surace}, {Teplitz}, {Yun}, \&
  {Wilson}}]{elbaz11}
{Elbaz}, D., {Dickinson}, M., {Hwang}, H.~S., {et~al.} 2011, \aap, 533, A119,
  \dodoi{10.1051/0004-6361/201117239}

\bibitem[{{Estrada-Carpenter} {et~al.}(2020){Estrada-Carpenter}, {Papovich},
  {Momcheva}, {Brammer}, {Simons}, {Bridge}, {Cleri}, {Ferguson},
  {Finkelstein}, {Giavalisco}, {Jung}, {Matharu}, {Trump}, \&
  {Weiner}}]{estrada_carpenter20}
{Estrada-Carpenter}, V., {Papovich}, C., {Momcheva}, I., {et~al.} 2020, arXiv
  e-prints, arXiv:2005.12289.
\newblock \doarXiv{2005.12289}

\bibitem[{{Flores Vel{\'a}zquez} {et~al.}(2021){Flores Vel{\'a}zquez},
  {Gurvich}, {Faucher-Gigu{\`e}re}, {Bullock}, {Starkenburg}, {Moreno},
  {Lazar}, {Mercado}, {Stern}, {Sparre}, {Hayward}, {Wetzel}, \&
  {El-Badry}}]{velazquez21}
{Flores Vel{\'a}zquez}, J.~A., {Gurvich}, A.~B., {Faucher-Gigu{\`e}re}, C.-A.,
  {et~al.} 2021, \mnras, 501, 4812, \dodoi{10.1093/mnras/staa3893}

\bibitem[{{Freundlich} {et~al.}(2019){Freundlich}, {Combes}, {Tacconi},
  {Genzel}, {Garcia-Burillo}, {Neri}, {Contini}, {Bolatto}, {Lilly},
  {Salom{\'e}}, {Bicalho}, {Boissier}, {Boone}, {Bouch{\'e}}, {Bournaud},
  {Burkert}, {Carollo}, {Cooper}, {Cox}, {Feruglio}, {F{\"o}rster Schreiber},
  {Juneau}, {Lippa}, {Lutz}, {Naab}, {Renzini}, {Saintonge}, {Sternberg},
  {Walter}, {Weiner}, {Wei{\ss}}, \& {Wuyts}}]{freundlich19}
{Freundlich}, J., {Combes}, F., {Tacconi}, L.~J., {et~al.} 2019, \aap, 622,
  A105, \dodoi{10.1051/0004-6361/201732223}

\bibitem[{{Georgakakis} {et~al.}(2017){Georgakakis}, {Aird}, {Schulze},
  {Dwelly}, {Salvato}, {Nandra}, {Merloni}, \& {Schneider}}]{georgakakis17}
{Georgakakis}, A., {Aird}, J., {Schulze}, A., {et~al.} 2017, \mnras, 471, 1976,
  \dodoi{10.1093/mnras/stx1602}

\bibitem[{{Greene} {et~al.}(2020){Greene}, {Strader}, \& {Ho}}]{greene20}
{Greene}, J.~E., {Strader}, J., \& {Ho}, L.~C. 2020, \araa, 58, 257,
  \dodoi{10.1146/annurev-astro-032620-021835}

\bibitem[{{Grogin} {et~al.}(2011){Grogin}, {Kocevski}, {Faber}, {Ferguson},
  {Koekemoer}, {Riess}, {Acquaviva}, {Alexander}, {Almaini}, {Ashby}, {Barden},
  {Bell}, {Bournaud}, {Brown}, {Caputi}, {Casertano}, {Cassata}, {Castellano},
  {Challis}, {Chary}, {Cheung}, {Cirasuolo}, {Conselice}, {Roshan Cooray},
  {Croton}, {Daddi}, {Dahlen}, {Dav{\'e}}, {de Mello}, {Dekel}, {Dickinson},
  {Dolch}, {Donley}, {Dunlop}, {Dutton}, {Elbaz}, {Fazio}, {Filippenko},
  {Finkelstein}, {Fontana}, {Gardner}, {Garnavich}, {Gawiser}, {Giavalisco},
  {Grazian}, {Guo}, {Hathi}, {H{\"a}ussler}, {Hopkins}, {Huang}, {Huang},
  {Jha}, {Kartaltepe}, {Kirshner}, {Koo}, {Lai}, {Lee}, {Li}, {Lotz}, {Lucas},
  {Madau}, {McCarthy}, {McGrath}, {McIntosh}, {McLure}, {Mobasher},
  {Moustakas}, {Mozena}, {Nandra}, {Newman}, {Niemi}, {Noeske}, {Papovich},
  {Pentericci}, {Pope}, {Primack}, {Rajan}, {Ravindranath}, {Reddy}, {Renzini},
  {Rix}, {Robaina}, {Rodney}, {Rosario}, {Rosati}, {Salimbeni}, {Scarlata},
  {Siana}, {Simard}, {Smidt}, {Somerville}, {Spinrad}, {Straughn}, {Strolger},
  {Telford}, {Teplitz}, {Trump}, {van der Wel}, {Villforth}, {Wechsler},
  {Weiner}, {Wiklind}, {Wild}, {Wilson}, {Wuyts}, {Yan}, \& {Yun}}]{grogin11}
{Grogin}, N.~A., {Kocevski}, D.~D., {Faber}, S.~M., {et~al.} 2011, \apjs, 197,
  35, \dodoi{10.1088/0067-0049/197/2/35}

\bibitem[{{Guo} {et~al.}(2013){Guo}, {Ferguson}, {Giavalisco}, {Barro},
  {Willner}, {Ashby}, {Dahlen}, {Donley}, {Faber}, {Fontana}, {Galametz},
  {Grazian}, {Huang}, {Kocevski}, {Koekemoer}, {Koo}, {McGrath}, {Peth},
  {Salvato}, {Wuyts}, {Castellano}, {Cooray}, {Dickinson}, {Dunlop}, {Fazio},
  {Gardner}, {Gawiser}, {Grogin}, {Hathi}, {Hsu}, {Lee}, {Lucas}, {Mobasher},
  {Nandra}, {Newman}, \& {van der Wel}}]{guo13}
{Guo}, Y., {Ferguson}, H.~C., {Giavalisco}, M., {et~al.} 2013, \apjs, 207, 24,
  \dodoi{10.1088/0067-0049/207/2/24}

\bibitem[{{Harrison}(2017)}]{harrison17}
{Harrison}, C.~M. 2017, Nature Astronomy, 1, 0165,
  \dodoi{10.1038/s41550-017-0165}

\bibitem[{{Hickox} {et~al.}(2014){Hickox}, {Mullaney}, {Alexander}, {Chen},
  {Civano}, {Goulding}, \& {Hainline}}]{hickox14}
{Hickox}, R.~C., {Mullaney}, J.~R., {Alexander}, D.~M., {et~al.} 2014, \apj,
  782, 9, \dodoi{10.1088/0004-637X/782/1/9}

\bibitem[{{Hodge} {et~al.}(2016){Hodge}, {Swinbank}, {Simpson}, {Smail},
  {Walter}, {Alexander}, {Bertoldi}, {Biggs}, {Brandt}, {Chapman}, {Chen},
  {Coppin}, {Cox}, {Dannerbauer}, {Edge}, {Greve}, {Ivison}, {Karim},
  {Knudsen}, {Menten}, {Rix}, {Schinnerer}, {Wardlow}, {Weiss}, \& {van der
  Werf}}]{hodge16}
{Hodge}, J.~A., {Swinbank}, A.~M., {Simpson}, J.~M., {et~al.} 2016, \apj, 833,
  103, \dodoi{10.3847/1538-4357/833/1/103}

\bibitem[{{Hopkins} {et~al.}(2022){Hopkins}, {Wellons},
  {Angl{\'e}s-Alc{\'a}zar}, {Faucher-Gigu{\`e}re}, \& {Grudi{\'c}}}]{hopkins22}
{Hopkins}, P.~F., {Wellons}, S., {Angl{\'e}s-Alc{\'a}zar}, D.,
  {Faucher-Gigu{\`e}re}, C.-A., \& {Grudi{\'c}}, M.~Y. 2022, \mnras, 510, 630,
  \dodoi{10.1093/mnras/stab3458}

\bibitem[{{Huertas-Company} {et~al.}(2015){Huertas-Company}, {Gravet},
  {Cabrera-Vives}, {P{\'e}rez-Gonz{\'a}lez}, {Kartaltepe}, {Barro}, {Bernardi},
  {Mei}, {Shankar}, {Dimauro}, {Bell}, {Kocevski}, {Koo}, {Faber}, \&
  {Mcintosh}}]{huertas_company15}
{Huertas-Company}, M., {Gravet}, R., {Cabrera-Vives}, G., {et~al.} 2015, \apjs,
  221, 8, \dodoi{10.1088/0067-0049/221/1/8}

\bibitem[{{Huertas-Company} {et~al.}(2016){Huertas-Company}, {Bernardi},
  {P{\'e}rez-Gonz{\'a}lez}, {Ashby}, {Barro}, {Conselice}, {Daddi}, {Dekel},
  {Dimauro}, {Faber}, {Grogin}, {Kartaltepe}, {Kocevski}, {Koekemoer}, {Koo},
  {Mei}, \& {Shankar}}]{huertas_company16}
{Huertas-Company}, M., {Bernardi}, M., {P{\'e}rez-Gonz{\'a}lez}, P.~G.,
  {et~al.} 2016, \mnras, 462, 4495, \dodoi{10.1093/mnras/stw1866}

\bibitem[{{Huertas-Company} {et~al.}(2018){Huertas-Company}, {Primack},
  {Dekel}, {Koo}, {Lapiner}, {Ceverino}, {Simons}, {Snyder}, {Bernardi},
  {Chen}, {Dom{\'{\i}}nguez-S{\'a}nchez}, {Lee}, {Margalef-Bentabol}, \&
  {Tuccillo}}]{huertas_company18}
{Huertas-Company}, M., {Primack}, J.~R., {Dekel}, A., {et~al.} 2018, \apj, 858,
  114, \dodoi{10.3847/1538-4357/aabfed}

\bibitem[{{Ivison} {et~al.}(2007){Ivison}, {Greve}, {Dunlop}, {Peacock},
  {Egami}, {Smail}, {Ibar}, {van Kampen}, {Aretxaga}, {Babbedge}, {Biggs},
  {Blain}, {Chapman}, {Clements}, {Coppin}, {Farrah}, {Halpern}, {Hughes},
  {Jarvis}, {Jenness}, {Jones}, {Mortier}, {Oliver}, {Papovich},
  {P{\'e}rez-Gonz{\'a}lez}, {Pope}, {Rawlings}, {Rieke}, {Rowan-Robinson},
  {Savage}, {Scott}, {Seigar}, {Serjeant}, {Simpson}, {Stevens}, {Vaccari},
  {Wagg}, \& {Willott}}]{ivison07}
{Ivison}, R.~J., {Greve}, T.~R., {Dunlop}, J.~S., {et~al.} 2007, \mnras, 380,
  199, \dodoi{10.1111/j.1365-2966.2007.12044.x}

\bibitem[{{Jahnke} \& {Macci{\`o}}(2011)}]{jahnke11}
{Jahnke}, K., \& {Macci{\`o}}, A.~V. 2011, \apj, 734, 92,
  \dodoi{10.1088/0004-637X/734/2/92}

\bibitem[{{Kennicutt}(1998)}]{kennicutt98}
{Kennicutt}, Jr., R.~C. 1998, \apj, 498, 541, \dodoi{10.1086/305588}

\bibitem[{{King} \& {Pounds}(2015)}]{king15}
{King}, A., \& {Pounds}, K. 2015, \araa, 53, 115,
  \dodoi{10.1146/annurev-astro-082214-122316}

\bibitem[{{Kirkpatrick} {et~al.}(2019){Kirkpatrick}, {Sharon}, {Keller}, \&
  {Pope}}]{kirkpatrick19}
{Kirkpatrick}, A., {Sharon}, C., {Keller}, E., \& {Pope}, A. 2019, \apj, 879,
  41, \dodoi{10.3847/1538-4357/ab223a}

\bibitem[{{Kocevski} {et~al.}(2017){Kocevski}, {Barro}, {Faber}, {Dekel},
  {Somerville}, {Young}, {Williams}, {McIntosh}, {Georgakakis}, {Hasinger},
  {Nandra}, {Civano}, {Alexander}, {Almaini}, {Conselice}, {Donley},
  {Ferguson}, {Giavalisco}, {Grogin}, {Hathi}, {Hawkins}, {Koekemoer}, {Koo},
  {McGrath}, {Mobasher}, {P{\'e}rez Gonz{\'a}lez}, {Pforr}, {Primack},
  {Santini}, {Stefanon}, {Trump}, {van der Wel}, {Wuyts}, \&
  {Yan}}]{kocevski17}
{Kocevski}, D.~D., {Barro}, G., {Faber}, S.~M., {et~al.} 2017, \apj, 846, 112,
  \dodoi{10.3847/1538-4357/aa8566}

\bibitem[{{Koekemoer} {et~al.}(2011){Koekemoer}, {Faber}, {Ferguson}, {Grogin},
  {Kocevski}, {Koo}, {Lai}, {Lotz}, {Lucas}, {McGrath}, {Ogaz}, {Rajan},
  {Riess}, {Rodney}, {Strolger}, {Casertano}, {Castellano}, {Dahlen},
  {Dickinson}, {Dolch}, {Fontana}, {Giavalisco}, {Grazian}, {Guo}, {Hathi},
  {Huang}, {van der Wel}, {Yan}, {Acquaviva}, {Alexander}, {Almaini}, {Ashby},
  {Barden}, {Bell}, {Bournaud}, {Brown}, {Caputi}, {Cassata}, {Challis},
  {Chary}, {Cheung}, {Cirasuolo}, {Conselice}, {Roshan Cooray}, {Croton},
  {Daddi}, {Dav{\'e}}, {de Mello}, {de Ravel}, {Dekel}, {Donley}, {Dunlop},
  {Dutton}, {Elbaz}, {Fazio}, {Filippenko}, {Finkelstein}, {Frazer}, {Gardner},
  {Garnavich}, {Gawiser}, {Gruetzbauch}, {Hartley}, {H{\"a}ussler},
  {Herrington}, {Hopkins}, {Huang}, {Jha}, {Johnson}, {Kartaltepe},
  {Khostovan}, {Kirshner}, {Lani}, {Lee}, {Li}, {Madau}, {McCarthy},
  {McIntosh}, {McLure}, {McPartland}, {Mobasher}, {Moreira}, {Mortlock},
  {Moustakas}, {Mozena}, {Nandra}, {Newman}, {Nielsen}, {Niemi}, {Noeske},
  {Papovich}, {Pentericci}, {Pope}, {Primack}, {Ravindranath}, {Reddy},
  {Renzini}, {Rix}, {Robaina}, {Rosario}, {Rosati}, {Salimbeni}, {Scarlata},
  {Siana}, {Simard}, {Smidt}, {Snyder}, {Somerville}, {Spinrad}, {Straughn},
  {Telford}, {Teplitz}, {Trump}, {Vargas}, {Villforth}, {Wagner}, {Wandro},
  {Wechsler}, {Weiner}, {Wiklind}, {Wild}, {Wilson}, {Wuyts}, \&
  {Yun}}]{koekemoer11}
{Koekemoer}, A.~M., {Faber}, S.~M., {Ferguson}, H.~C., {et~al.} 2011, \apjs,
  197, 36, \dodoi{10.1088/0067-0049/197/2/36}

\bibitem[{{Kormendy} \& {Ho}(2013)}]{kormendy13}
{Kormendy}, J., \& {Ho}, L.~C. 2013, \araa, 51, 511,
  \dodoi{10.1146/annurev-astro-082708-101811}

\bibitem[{{Lamperti} {et~al.}(2020){Lamperti}, {Saintonge}, {Koss}, {Viti},
  {Wilson}, {He}, {Shimizu}, {Greve}, {Mushotzky}, {Treister}, {Kramer},
  {Sanders}, {Schawinski}, \& {Tacconi}}]{lamperti20}
{Lamperti}, I., {Saintonge}, A., {Koss}, M., {et~al.} 2020, \apj, 889, 103,
  \dodoi{10.3847/1538-4357/ab6221}

\bibitem[{{Luo} {et~al.}(2017){Luo}, {Brandt}, {Xue}, {Lehmer}, {Alexander},
  {Bauer}, {Vito}, {Yang}, {Basu-Zych}, {Comastri}, {Gilli}, {Gu},
  {Hornschemeier}, {Koekemoer}, {Liu}, {Mainieri}, {Paolillo}, {Ranalli},
  {Rosati}, {Schneider}, {Shemmer}, {Smail}, {Sun}, {Tozzi}, {Vignali}, \&
  {Wang}}]{luo17}
{Luo}, B., {Brandt}, W.~N., {Xue}, Y.~Q., {et~al.} 2017, \apjs, 228, 2,
  \dodoi{10.3847/1538-4365/228/1/2}

\bibitem[{{Lutz} {et~al.}(2011){Lutz}, {Poglitsch}, {Altieri}, {Andreani},
  {Aussel}, {Berta}, {Bongiovanni}, {Brisbin}, {Cava}, {Cepa}, {Cimatti},
  {Daddi}, {Dominguez-Sanchez}, {Elbaz}, {F{\"o}rster Schreiber}, {Genzel},
  {Grazian}, {Gruppioni}, {Harwit}, {Le Floc'h}, {Magdis}, {Magnelli},
  {Maiolino}, {Nordon}, {P{\'e}rez Garc{\'{\i}}a}, {Popesso}, {Pozzi},
  {Riguccini}, {Rodighiero}, {Saintonge}, {Sanchez Portal}, {Santini}, {Shao},
  {Sturm}, {Tacconi}, {Valtchanov}, {Wetzstein}, \& {Wieprecht}}]{lutz11}
{Lutz}, D., {Poglitsch}, A., {Altieri}, B., {et~al.} 2011, \aap, 532, A90,
  \dodoi{10.1051/0004-6361/201117107}

\bibitem[{{McMullin} {et~al.}(2007){McMullin}, {Waters}, {Schiebel}, {Young},
  \& {Golap}}]{mcmullin07}
{McMullin}, J.~P., {Waters}, B., {Schiebel}, D., {Young}, W., \& {Golap}, K.
  2007, in Astronomical Society of the Pacific Conference Series, Vol. 376,
  Astronomical Data Analysis Software and Systems XVI, ed. R.~A. {Shaw},
  F.~{Hill}, \& D.~J. {Bell}, 127

\bibitem[{{Mountrichas} {et~al.}(2021){Mountrichas}, {Buat}, {Yang}, {Boquien},
  {Burgarella}, {Ciesla}, {Malek}, \& {Shirley}}]{mountrichas21b}
{Mountrichas}, G., {Buat}, V., {Yang}, G., {et~al.} 2021, arXiv e-prints,
  arXiv:2106.10678.
\newblock \doarXiv{2106.10678}

\bibitem[{{Ni} {et~al.}(2019){Ni}, {Yang}, {Brandt}, {Alexander}, {Chen},
  {Luo}, {Vito}, \& {Xue}}]{ni19}
{Ni}, Q., {Yang}, G., {Brandt}, W.~N., {et~al.} 2019, \mnras, 490, 1135,
  \dodoi{10.1093/mnras/stz2623}

\bibitem[{{Ni} {et~al.}(2021){Ni}, {Brandt}, {Yang}, {Leja}, {Chen}, {Luo},
  {Matharu}, {Sun}, {Vito}, {Xue}, \& {Zhang}}]{ni21}
{Ni}, Q., {Brandt}, W.~N., {Yang}, G., {et~al.} 2021, \mnras, 500, 4989,
  \dodoi{10.1093/mnras/staa3514}

\bibitem[{{Papovich} {et~al.}(2005){Papovich}, {Dickinson}, {Giavalisco},
  {Conselic e}, \& {Ferguson}}]{papovich05}
{Papovich}, C., {Dickinson}, M., {Giavalisco}, M., {Conselic e}, C.~J., \&
  {Ferguson}, H.~C. 2005, \apj, 631, 101, \dodoi{10.1086/429120}

\bibitem[{{Peng}(2007)}]{peng07}
{Peng}, C.~Y. 2007, \apj, 671, 1098, \dodoi{10.1086/522774}

\bibitem[{{Ramos Padilla} {et~al.}(2021){Ramos Padilla}, {Wang}, {Ma{\l}ek},
  {Efstathiou}, \& {Yang}}]{ramos21}
{Ramos Padilla}, A.~F., {Wang}, L., {Ma{\l}ek}, K.~n., {Efstathiou}, A., \&
  {Yang}, G. 2021, arXiv e-prints, arXiv:2108.10899.
\newblock \doarXiv{2108.10899}

\bibitem[{{Roehlly} {et~al.}(2014){Roehlly}, {Burgarella}, {Buat}, {Boquien},
  {Ciesla}, \& {Heinis}}]{roehlly14}
{Roehlly}, Y., {Burgarella}, D., {Buat}, V., {et~al.} 2014, in Astronomical
  Society of the Pacific Conference Series, Vol. 485, Astronomical Data
  Analysis Software and Systems XXIII, ed. N.~{Manset} \& P.~{Forshay}, 347.
\newblock \doarXiv{1309.6366}

\bibitem[{{Saglia} {et~al.}(2016){Saglia}, {Opitsch}, {Erwin}, {Thomas},
  {Beifiori}, {Fabricius}, {Mazzalay}, {Nowak}, {Rusli}, \&
  {Bender}}]{saglia16}
{Saglia}, R.~P., {Opitsch}, M., {Erwin}, P., {et~al.} 2016, \apj, 818, 47,
  \dodoi{10.3847/0004-637X/818/1/47}

\bibitem[{{Saintonge} {et~al.}(2017){Saintonge}, {Catinella}, {Tacconi},
  {Kauffmann}, {Genzel}, {Cortese}, {Dav{\'e}}, {Fletcher},
  {Graci{\'a}-Carpio}, {Kramer}, {Heckman}, {Janowiecki}, {Lutz}, {Rosario},
  {Schiminovich}, {Schuster}, {Wang}, {Wuyts}, {Borthakur}, {Lamperti}, \&
  {Roberts-Borsani}}]{saintonge17}
{Saintonge}, A., {Catinella}, B., {Tacconi}, L.~J., {et~al.} 2017, \apjs, 233,
  22, \dodoi{10.3847/1538-4365/aa97e0}

\bibitem[{{Salim} {et~al.}(2007){Salim}, {Rich}, {Charlot}, {Brinchmann},
  {Johnson}, {Schiminovich}, {Seibert}, {Mallery}, {Heckman}, {Forster},
  {Friedman}, {Martin}, {Morrissey}, {Neff}, {Small}, {Wyder}, {Bianchi},
  {Donas}, {Lee}, {Madore}, {Milliard}, {Szalay}, {Welsh}, \& {Yi}}]{salim07}
{Salim}, S., {Rich}, R.~M., {Charlot}, S., {et~al.} 2007, \apjs, 173, 267,
  \dodoi{10.1086/519218}

\bibitem[{{Scoville} {et~al.}(2016){Scoville}, {Sheth}, {Aussel}, {Vanden
  Bout}, {Capak}, {Bongiorno}, {Casey}, {Murchikova}, {Koda},
  {{\'A}lvarez-M{\'a}rquez}, {Lee}, {Laigle}, {McCracken}, {Ilbert}, {Pope},
  {Sanders}, {Chu}, {Toft}, {Ivison}, \& {Manohar}}]{scoville16}
{Scoville}, N., {Sheth}, K., {Aussel}, H., {et~al.} 2016, \apj, 820, 83,
  \dodoi{10.3847/0004-637X/820/2/83}

\bibitem[{{Scoville} {et~al.}(2017){Scoville}, {Lee}, {Vanden Bout},
  {Diaz-Santos}, {Sanders}, {Darvish}, {Bongiorno}, {Casey}, {Murchikova},
  {Koda}, {Capak}, {Vlahakis}, {Ilbert}, {Sheth}, {Morokuma-Matsui}, {Ivison},
  {Aussel}, {Laigle}, {McCracken}, {Armus}, {Pope}, {Toft}, \&
  {Masters}}]{scoville17}
{Scoville}, N., {Lee}, N., {Vanden Bout}, P., {et~al.} 2017, \apj, 837, 150,
  \dodoi{10.3847/1538-4357/aa61a0}

\bibitem[{{Shangguan} {et~al.}(2020){Shangguan}, {Ho}, {Bauer}, {Wang}, \&
  {Treister}}]{shangguan20}
{Shangguan}, J., {Ho}, L.~C., {Bauer}, F.~E., {Wang}, R., \& {Treister}, E.
  2020, \apj, 899, 112, \dodoi{10.3847/1538-4357/aba8a1}

\bibitem[{{Solomon} \& {Vanden Bout}(2005)}]{solomon05}
{Solomon}, P.~M., \& {Vanden Bout}, P.~A. 2005, \araa, 43, 677,
  \dodoi{10.1146/annurev.astro.43.051804.102221}

\bibitem[{{Stalevski} {et~al.}(2012){Stalevski}, {Fritz}, {Baes}, {Nakos}, \&
  {Popovi{\'c}}}]{stalevski12}
{Stalevski}, M., {Fritz}, J., {Baes}, M., {Nakos}, T., \& {Popovi{\'c}},
  L.~{\v{C}}. 2012, \mnras, 420, 2756, \dodoi{10.1111/j.1365-2966.2011.19775.x}

\bibitem[{{Stalevski} {et~al.}(2016){Stalevski}, {Ricci}, {Ueda}, {Lira},
  {Fritz}, \& {Baes}}]{stalevski16}
{Stalevski}, M., {Ricci}, C., {Ueda}, Y., {et~al.} 2016, \mnras, 458, 2288,
  \dodoi{10.1093/mnras/stw444}

\bibitem[{{Suh} {et~al.}(2015){Suh}, {Hasinger}, {Steinhardt}, {Silverman}, \&
  {Schramm}}]{suh15}
{Suh}, H., {Hasinger}, G., {Steinhardt}, C., {Silverman}, J.~D., \& {Schramm},
  M. 2015, \apj, 815, 129, \dodoi{10.1088/0004-637X/815/2/129}

\bibitem[{{Vanzella} {et~al.}(2008){Vanzella}, {Cristiani}, {Dickinson},
  {Giavalisco}, {Kuntschner}, {Haase}, {Nonino}, {Rosati}, {Cesarsky},
  {Ferguson}, {Fosbury}, {Grazian}, {Moustakas}, {Rettura}, {Popesso},
  {Renzini}, {Stern}, \& {GOODS Team}}]{vanzella08}
{Vanzella}, E., {Cristiani}, S., {Dickinson}, M., {et~al.} 2008, \aap, 478, 83,
  \dodoi{10.1051/0004-6361:20078332}

\bibitem[{{Whitaker} {et~al.}(2012){Whitaker}, {van Dokkum}, {Brammer}, \&
  {Franx}}]{whitaker12}
{Whitaker}, K.~E., {van Dokkum}, P.~G., {Brammer}, G., \& {Franx}, M. 2012,
  \apjl, 754, L29, \dodoi{10.1088/2041-8205/754/2/L29}

\bibitem[{{Yang} {et~al.}(2019){Yang}, {Brandt}, {Alexander}, {Chen}, {Ni},
  {Vito}, \& {Zhu}}]{yang19}
{Yang}, G., {Brandt}, W.~N., {Alexander}, D.~M., {et~al.} 2019, \mnras, 485,
  3721, \dodoi{10.1093/mnras/stz611}

\bibitem[{{Yang} {et~al.}(2021){Yang}, {Estrada-Carpenter}, {Papovich}, {Vit
  o}, {Walsh}, {Yao}, \& {Yuan}}]{yang21b}
{Yang}, G., {Estrada-Carpenter}, V., {Papovich}, C., {et~al.} 2021, \apj, 921,
  170, \dodoi{10.3847/1538-4357/ac2233}

\bibitem[{{Yang} {et~al.}(2017){Yang}, {Chen}, {Vito}, {Brandt}, {Alexander},
  {Luo}, {Sun}, {Xue}, {Bauer}, {Koekemoer}, {Lehmer}, {Liu}, {Schneider},
  {Shemmer}, {Trump}, {Vignali}, \& {Wang}}]{yang17}
{Yang}, G., {Chen}, C.-T.~J., {Vito}, F., {et~al.} 2017, \apj, 842, 72,
  \dodoi{10.3847/1538-4357/aa7564}

\bibitem[{{Yang} {et~al.}(2018){Yang}, {Brandt}, {Vito}, {Chen}, {Trump},
  {Luo}, {Sun}, {Xue}, {Koekemoer}, {Schneider}, {Vignali}, \& {Wang}}]{yang18}
{Yang}, G., {Brandt}, W.~N., {Vito}, F., {et~al.} 2018, \mnras, 475, 1887,
  \dodoi{10.1093/mnras/stx2805}

\bibitem[{{Yang} {et~al.}(2020){Yang}, {Boquien}, {Buat}, {Burgarella},
  {Ciesla}, {Duras}, {Stalevski}, {Brandt}, \& {Papovich}}]{yang20}
{Yang}, G., {Boquien}, M., {Buat}, V., {et~al.} 2020, \mnras, 491, 740,
  \dodoi{10.1093/mnras/stz3001}

\bibitem[{{Yang} {et~al.}(2022){Yang}, {Boquien}, {Brandt}, {Buat},
  {Burgarella}, {Ciesla}, {Lehmer}, {Ma{\l}ek}, {Mountrichas}, {Papovich},
  {Pons}, {Stalevski}, {Theul{\'e}}, \& {Zhu}}]{yang22}
{Yang}, G., {Boquien}, M., {Brandt}, W.~N., {et~al.} 2022, \apj, 927, 192,
  \dodoi{10.3847/1538-4357/ac4971}

\end{thebibliography}
\bibliographystyle{aasjournal}


\end{CJK*}
\end{document}